\documentclass[pra,reprint,superscriptaddress,longbibliography]{revtex4-2}

\usepackage{array}
\usepackage{natbib}
\usepackage[section]{placeins}
\usepackage{pgfplots,amsfonts}
\usepackage[breaklinks]{hyperref}
\usepackage{amssymb}
\usepackage{amsmath,amsthm,bm}
\usepackage{amsfonts}
\usepackage{cleveref}
\usepackage{verbatim}
\usepackage{mathtools}
\usepackage{tikz}
\hypersetup{citecolor=red,colorlinks=true,urlcolor=blue}
\usepackage{algorithm,setspace}
\usepackage{algpseudocode}
\usetikzlibrary{angles, quotes}
\usetikzlibrary{positioning}

\usepackage[utf8]{inputenc} 
\DeclareUnicodeCharacter{0308}{\"{}} 
\DeclareUnicodeCharacter{03C0}{$\pi$} 

\usepackage[normalem]{ulem}
\usepackage{color}
\usepackage{xcolor}

\newcolumntype{P}[1]{>{\centering\arraybackslash}p{#1}}

\definecolor{Ugreen}{HTML}{198a11}

\begin{document}

\title{Optimally Weighted Constrained Shadow Tomography for Reduced Density Matrices}
\author{Yuchen Wang}
\affiliation{Department of Chemistry and The James Franck Institute, The University of Chicago, Chicago, IL 60637 USA}
\author{Irma Avdic}
\affiliation{IBM Research, Chicago, IL 60606 USA}
\author{Michael Rose}
\affiliation{Department of Chemistry and The James Franck Institute, The University of Chicago, Chicago, IL 60637 USA}
\author{Lillian I. Payne}
\affiliation{Department of Chemistry and The James Franck Institute, The University of Chicago, Chicago, IL 60637 USA}
\author{Noah Garrett}
\affiliation{Department of Chemistry and The James Franck Institute, The University of Chicago, Chicago, IL 60637 USA}
\author{Madhubani Mukherjee}
\affiliation{Department of Chemistry and The James Franck Institute, The University of Chicago, Chicago, IL 60637 USA}
\author{Leah P. Weisburn}
\affiliation{Department of Chemistry and The James Franck Institute, The University of Chicago, Chicago, IL 60637 USA}
\author{Nate Earnest-Noble}
\affiliation{IBM Research, Yorktown Heights, NY 10598 USA}
\author{Kevin J. Sung}
\email{kevinsung@ibm.com}
\affiliation{IBM Research, Yorktown Heights, NY 10598 USA}
\author{David A. Mazziotti}
\email{damazz@uchicago.edu}
\affiliation{Department of Chemistry and The James Franck Institute, The University of Chicago, Chicago, IL 60637 USA}

\date{Submitted October 26, 2026}

\begin{abstract}
Accurate measurement of reduced density matrices (RDMs) is central to quantum simulations of many-fermion systems, but the required measurement resources remain a significant challenge. Constrained shadow tomography (CST) combines randomized measurements with $N$-representability conditions to reconstruct physically valid RDMs from limited measurement data. Here we improve CST by orthogonalizing the measurement constraints and using the resulting singular values to determine the relative weight of each independent measurement direction according to how strongly it is sampled by the quantum computer. These optimally weighted constraints are incorporated into the semidefinite program through a least-squares objective implemented with a Schur-complement relaxation. We demonstrate the approach for hydrogen chains on a quantum simulator and for azobenzene on a quantum processor. In both cases, optimally weighted CST systematically improves the accuracy of the reconstructed 2-RDM without increasing the measurement budget, providing a more measurement-efficient approach to partial tomography for quantum simulations of correlated fermions.
\end{abstract}

\maketitle

\section{Introduction}

Quantum-state tomography provides a route to determining the properties of quantum states prepared on quantum computers, but conventional tomography requires a number of measurements that grows exponentially with system size~\cite{James2001,bonet2020nearly,Vogel1989,Hradil1997,Paris2004,Haah2017}. Shadow tomography and classical shadows provide an important alternative by using randomized measurements to estimate large collections of observables with substantially fewer measurements~\cite{Aaronson2020,Huang2020,Zhao2021,Elben2022,Aaronson2019,Elben2020,Struchalin2021,Badescu2021,Huang2021,Chen2021,Huggins2022,Hadfield2022,Koh2022,Bertoni2024,Sack2022,Seif2023,Hu2023,Levy2024}. Rather than reconstructing the complete many-particle wave function or density matrix, the measurements can therefore be directed toward the quantities needed for a given quantum simulation.

For many-fermion systems, reduced density matrices (RDMs) provide a particularly natural target for such partial tomography. The $p$-particle RDM contains the information required to evaluate all $p$-particle observables while growing only polynomially with the size of the one-particle basis. In particular, the two-particle reduced density matrix (2-RDM) determines the energy and other two-particle properties of an electronic system with pairwise interactions~\cite{Gidofalvi2007,Overy2014,OBrien2019}. Measurements of RDMs have consequently become an important component of quantum algorithms for electronic structure and many-body simulation~\cite{rubin2018application,motta2020qite,bauer2016embedding,obrien2022forces,smart2021cqe,wang2023excited,wang2025rlcqe,McArdle2020,HeadMarsden2021,Jiang2020,Smart2022}. More recently, fermionic shadow tomography has adapted the classical-shadow framework directly to fermionic observables and RDMs, providing an efficient route to their measurement from randomized orbital transformations~\cite{low2022classical,Wan2023,Huang2024,ZhaoMiyake2024,King2025,wang2025shadow}.

An important additional source of information is that an RDM arising from an $N$-fermion quantum state cannot be arbitrary. It must satisfy the $N$-representability conditions that encode its consistency with an underlying $N$-particle density matrix~\cite{Coleman1963,Kummer1967,Erdahl1978,ColemanYukalov2000}. In our previous work, we introduced constrained shadow tomography (CST), which combines the linear information obtained from randomized shadow measurements with these $N$-representability conditions in a variational semidefinite-programming calculation of the 2-RDM~\cite{Avdic2024,Avdic2024excited,avdic2026constrained,wang2026correlated}. In this way, the quantum measurements determine selected linear projections of the 2-RDM, while the $N$-representability conditions constrain the remaining degrees of freedom to the physically allowed region. The combination provides substantially more information than either the shadow measurements or the variational 2-RDM constraints~\cite{MazziottiErdahl2001,Nakata2001,Mazziotti2002,M2004,Zhao2004,Gidofalvi2005,Cances2006,gidofalvi2008activespace,Shenvi2010,Mazziotti2016,Alcoba2018} alone.

Here we introduce optimally weighted constrained shadow tomography. The measurement constraints generated by different randomized unitaries generally sample different independent directions of the 2-RDM space to different extents. We orthogonalize these constraints through a singular-value decomposition of the measurement matrix, for which the singular values quantify how strongly each independent direction has been sampled by the quantum measurements. These singular values then provide the natural relative weights of the measurement constraints in the reconstruction. We incorporate the weighted constraints into the semidefinite program through a least-squares objective implemented by a Schur-complement relaxation. The resulting formulation extracts more information from the same set of quantum measurements and systematically improves the accuracy of the reconstructed 2-RDM without increasing the measurement budget.

We demonstrate the approach for linear hydrogen chains on a quantum simulator and for azobenzene on a quantum processor. The results show that optimally weighted CST yields more accurate 2-RDMs at fixed measurement cost and provides a systematic route for combining fermionic shadow measurements with the physical information contained in $N$-representability.

\section{Theory}

\subsection{Constrained Shadow Tomography and $N$-Representability}

For a many-electron quantum system with a Hamiltonian containing no more than pairwise interactions, the energy can be expressed exactly as a linear functional of the two-electron reduced density matrix (2-RDM),
\begin{equation}
E=\mathrm{Tr}\!\left({}^{2}K\,{}^{2}D\right),
\end{equation}
where ${}^{2}K$ is the reduced two-electron Hamiltonian and the elements of the 2-RDM are
\begin{equation}
{}^{2}D^{ij}_{kl}
=
\langle\Psi|
\hat a_i^\dagger \hat a_j^\dagger
\hat a_l \hat a_k
|\Psi\rangle .
\end{equation}
The 2-RDM likewise determines all one- and two-particle properties while requiring only a polynomial number of elements in the size of the one-particle basis. It is therefore a natural target for partial quantum-state tomography of many-electron systems.

A hierarchy of $N$-representability conditions constrains the 2-RDM to be consistent with an $N$-electron ensemble density matrix~\cite{M2007,M2012,M2012b,Garrod1964,MazziottiErdahl2001,Mazziotti2006,Mazziotti2023,Mazziotti.2026}. At the level of two-particle operators, the 2-positivity or DQG conditions require
\begin{equation}
{}^{2}D\succeq0,\qquad
{}^{2}Q\succeq0,\qquad
{}^{2}G\succeq0 ,
\end{equation}
where the matrix elements are
\begin{align}
{}^{2}D^{ij}_{kl}
&=
\langle\Psi|
\hat a_i^\dagger \hat a_j^\dagger
\hat a_l \hat a_k
|\Psi\rangle ,
\\
{}^{2}Q^{ij}_{kl}
&=
\langle\Psi|
\hat a_i \hat a_j
\hat a_l^\dagger \hat a_k^\dagger
|\Psi\rangle ,
\\
{}^{2}G^{ij}_{kl}
&=
\langle\Psi|
\hat a_i^\dagger \hat a_j
\hat a_l^\dagger \hat a_k
|\Psi\rangle .
\end{align}
The $D$, $Q$, and $G$ conditions ensure nonnegative probabilities for arbitrary two-particle, two-hole, and particle-hole distributions, respectively. Because ${}^{2}Q$ and ${}^{2}G$ are linear functions of ${}^{2}D$, these conditions provide direct physical constraints on the space of possible 2-RDMs.

Shadows of the 2-RDM can be obtained by applying randomized orbital transformations and measuring pair occupations in the rotated basis. For the $n$th random unitary transformation $U_n$, a measured shadow element can be written as
\begin{equation}
S_n^{pq}
=
\langle\Psi|
\hat U_n^\dagger
\hat a_p^\dagger \hat a_q^\dagger
\hat a_q \hat a_p
\hat U_n
|\Psi\rangle ,
\label{eq:shadow}
\end{equation}
which corresponds to a diagonal element of the 2-RDM in the randomly rotated orbital basis,
\begin{equation}
S_n^{pq}
=
\left[
(U_n\otimes U_n)\,
{}^{2}D\,
(U_n\otimes U_n)^T
\right]_{pq}^{pq}.
\end{equation}
Each shadow measurement therefore provides a linear projection of the 2-RDM.

In fermionic shadow tomography (FST), the measured projections are combined statistically to estimate the elements of the 2-RDM. Constrained shadow tomography (CST) instead incorporates the same measured projections directly into a variational determination of the 2-RDM subject to the $N$-representability conditions. The shadow measurements therefore provide information about selected directions of the 2-RDM, while the $N$-representability conditions constrain the remaining degrees of freedom to correspond to a physically admissible $N$-electron density matrix. By combining the measured information with these exact fermionic constraints, CST can obtain a more accurate 2-RDM from a limited set of quantum measurements than either the shadow estimates or the variational $N$-representability conditions alone.

The CST reconstruction can therefore be formulated as the semidefinite program
\begin{align}
\min_{{}^{2}D,{}^{2}\tilde D}\quad
&E[{}^{2}D]
+
\left\|{}^{2}D-{}^{2}\tilde{D}\right\|_{*}
\label{eq:CST}
\\
\text{subject to}\quad
&{}^{2}D \succeq 0,
\nonumber\\
&{}^{2}Q \succeq 0,
\nonumber\\
&{}^{2}G \succeq 0,
\nonumber\\
&{}^{2}Q=f_Q({}^{2}D),
\nonumber\\
&{}^{2}G=f_G({}^{2}D),
\nonumber\\
&S_n^{pq}
=
\left[
(U_n\otimes U_n)\,
{}^{2}\tilde D\,
(U_n\otimes U_n)^T
\right]_{pq}^{pq}.
\nonumber
\end{align}
Here ${}^{2}\tilde D$ is constrained by the measured shadows, while
${}^{2}D$ is constrained to satisfy the $N$-representability conditions.
The nuclear-norm term~\cite{Recht2010,Candes2009,Foley2012} couples the measured and physically admissible
2-RDMs, while the energy provides the variational criterion for selecting
${}^{2}D$. Further details of the bi-objective formulation are given in
Refs.~\cite{avdic2026constrained,wang2026correlated}.

In the original CST formulation, the measured shadow constraints enter the reconstruction without distinguishing how strongly different independent directions of the 2-RDM space have been sampled. The randomized measurements, however, generally contain unequal amounts of information about these directions. In the present work, we make this information explicit by transforming the measurement constraints to an orthogonal basis and using the resulting singular values to determine their optimal relative weights. This weighted formulation allows the information contained in a fixed set of quantum measurements to be used more effectively in the reconstruction of the 2-RDM.

\subsection{Orthogonalization and Optimal Weighting of Measurement Constraints}
\label{sec:ortho}

The shadow measurements do not, in general, sample all directions of the
2-RDM space equally. Some measurement directions are sampled more strongly
than others. To determine how strongly each independent direction has been
sampled, we first express the shadow measurements as a set of linear
constraints on the 2-RDM.

Let $\hat M_k$ denote the two-particle measurement operator associated
with the $k$th measured shadow value $v_k$. Each measurement provides
the linear constraint
\begin{equation}
v_k
=
\langle\Psi|\hat M_k|\Psi\rangle
=
\mathrm{Tr}\!\left(M_k\,{}^2D\right),
\end{equation}
where $M_k$ is the matrix representation of $\hat M_k$ in the
two-particle space.

Vectorizing the measurement matrices and collecting them as columns
defines the measurement matrix
\begin{equation}
A=
\begin{bmatrix}
\mathrm{vec}(M_1) &
\mathrm{vec}(M_2) &
\cdots &
\mathrm{vec}(M_{N_c})
\end{bmatrix}.
\end{equation}
The complete set of measured constraints can then be written as
\begin{equation}
A^T\mathrm{vec}({}^2D)=\mathbf{v},
\label{eq:linear_system}
\end{equation}
where $\mathbf{v}$ contains the corresponding measured expectation
values.

The information contained in these measurements is exposed by the
singular-value decomposition
\begin{equation}
A=U\Sigma V^T.
\end{equation}
Each part of this decomposition has a direct interpretation in the
tomography. The right singular vectors in $V$ identify independent
linear combinations of the original shadow measurements. The
corresponding left singular vectors in $U$ define orthonormal directions
in the space of two-particle operators that are probed by those
measurement combinations. Finally, the singular value $\sigma_i$
associated with each pair of singular vectors measures how strongly
that independent direction has been sampled by the complete set of
shadow measurements. Equivalently,
\begin{equation}
A V_i=\sigma_i U_i,
\end{equation}
showing that the $i$th combination of the original measurements,
specified by $V_i$, probes the normalized 2-RDM direction $U_i$ with
strength $\sigma_i$.

Substitution of the singular-value decomposition into
Eq.~\eqref{eq:linear_system} gives
\begin{equation}
U^T\mathrm{vec}({}^2D)
=
\Sigma^{-1}V^T\mathbf{v}.
\label{eq:transformed_system}
\end{equation}
The original shadow constraints are therefore replaced by measurements
along orthogonal directions of the 2-RDM space. The transformed
measurement values are
\begin{equation}
\tilde{\mathbf{v}}_{\mathrm{ortho}}
=
\Sigma^{-1}V^T\mathbf{v},
\end{equation}
with corresponding constraints
\begin{equation}
\mathrm{Tr}\!\left(
\tilde M_{i,\mathrm{ortho}}\,{}^2D
\right)
=
\tilde v_{i,\mathrm{ortho}}.
\end{equation}

The singular values provide the relative sampling weight of these
independent constraints. We define
\begin{equation}
w_i=\frac{\sigma_i}{\sigma_1},
\label{eq:weights}
\end{equation}
where $\sigma_1$ is the largest singular value.  Because the original measurement constraints are nonorthogonal, they do not directly reflect the relative sampling strengths of independent directions in the 2-RDM space. We correct this by orthogonalizing the measurement constraints and assigning suitable weights to each orthogonal direction as provided by the singular values.

\subsection{Weighted Least-Squares Semidefinite Program}
\label{sec:weighted_sdp}

The singular-value decomposition provides a relative weight $w_i$ for
each independent measurement direction. We incorporate these weights
into constrained shadow tomography through a weighted least-squares
objective. For each orthogonalized measurement constraint, we define a
residual $e_i$ according to
\begin{equation}
\mathrm{Tr}\!\left(
\tilde M_{i,\mathrm{ortho}}\,{}^2D
\right)
+e_i
=
\tilde v_{i,\mathrm{ortho}},
\qquad i=1,\ldots,r',
\label{eq:residual}
\end{equation}
where $r'$ is the number of retained independent measurement
directions. The measurement contribution to the objective is then
\begin{equation}
\sum_{i=1}^{r'} w_i^2 e_i^2 .
\label{eq:weighted_ls}
\end{equation}
Consequently, deviations from strongly sampled measurement directions
carry greater weight than deviations from directions that are only
weakly sampled.

Because the quadratic least-squares term cannot appear directly as a
linear semidefinite-programming objective, we introduce auxiliary
variables $f_i$ satisfying
\begin{equation}
f_i \ge e_i^2 .
\end{equation}
These inequalities can be imposed through the Schur-complement
condition
\begin{equation}
Z=
\begin{pmatrix}
I_{r'} & E\\
E & F
\end{pmatrix}
\succeq 0,
\label{eq:Zdef}
\end{equation}
where
\begin{equation}
E=\operatorname{diag}(e_1,\ldots,e_{r'}),
\qquad
F=\operatorname{diag}(f_1,\ldots,f_{r'}).
\end{equation}
The Schur complement of the identity block gives
\begin{equation}
F-E^2\succeq0,
\end{equation}
and hence $f_i\ge e_i^2$ for every measurement direction.

The optimally weighted CST reconstruction can therefore be written as
\begin{align}
\min_{{}^2D,\{e_i\},\{f_i\}}\quad
&E[{}^2D]
+
\lambda\sum_{i=1}^{r'}w_i^2 f_i
\label{eq:weighted_CST}
\\
\text{subject to}\quad
&{}^2D\succeq0,
\nonumber\\
&{}^2Q\succeq0,
\nonumber\\
&{}^2G\succeq0,
\nonumber\\
&{}^2Q=f_Q({}^2D),
\nonumber\\
&{}^2G=f_G({}^2D),
\nonumber\\
&\mathrm{Tr}({}^2D)=N(N-1),
\nonumber\\
&\mathrm{Tr}\!\left(
\tilde M_{i,\mathrm{ortho}}\,{}^2D
\right)+e_i
=
\tilde v_{i,\mathrm{ortho}},
\, i \in [1,r'],
\nonumber\\
&Z\succeq0 .
\nonumber
\end{align}
Here $\lambda$ controls the relative importance of the variational
energy and the information supplied by the quantum measurements. In
the limit of exact and sufficiently complete measurements, the
residuals vanish and the measured projections are recovered exactly.
For finite measurement data, the least-squares formulation allows the
2-RDM to reconcile the measured information with the
$N$-representability conditions, with each independent measurement
direction contributing according to how strongly it has been sampled.

\subsection{Computational Scaling}
\label{sec:scaling}

A direct singular-value decomposition of the measurement matrix
$A\in\mathbb{R}^{r^4\times N_c}$ would appear to require construction
of a large two-particle matrix. This cost can be avoided because each
two-particle measurement observable is generated from products of
one-particle orbital rotations.

Let $\phi_p^{(n)}$ denote orbital $p$ in measurement setting $n$, and
define the one-particle overlap matrix between settings $n$ and $m$ by
\begin{equation}
O^{nm}_{pp'}
=
\langle\phi_p^{(n)}|\phi_{p'}^{(m)}\rangle .
\end{equation}
The overlap between two pair-measurement operators can then be obtained
directly from these one-particle overlaps,
\begin{equation}
\left\langle
M_{n,pq},M_{m,p'q'}
\right\rangle
=
\left|
O^{nm}_{pp'}O^{nm}_{qq'}
-
O^{nm}_{pq'}O^{nm}_{qp'}
\right|^2 .
\end{equation}
Hence, the Gram matrix
\begin{equation}
G=A^T A
\end{equation}
can be constructed without explicitly forming the full measurement
matrix $A$.

Because $N_c=O(N_u r^2)$, construction of the Gram matrix requires
$O(N_u^2r^4)$ elements, and its direct eigendecomposition scales as
\begin{equation}
O(N_c^3)=O(N_u^3r^6).
\end{equation}
For the small number of randomized measurement settings used here, this
classical preprocessing remains tractable and preserves the $r^6$
scaling with orbital dimension for fixed $N_u$.

The factorized structure of the measurement map also permits
matrix-free applications of $A$ and $A^T$, suggesting that the cost of
the orthogonalization can be reduced further for larger measurement
sets. We leave the development of such structured decompositions to
future work.

\section{Results}

We apply the optimally weighted CST method to linear hydrogen chains~\cite{Suhai1994} on a
quantum simulator and to azobenzene on a quantum processor. The hydrogen
chains provide a controlled benchmark for examining the dependence of
the reconstruction on the number of randomized measurement settings,
the number of shots, and the system size. We use a uniform internuclear
separation of $1.0$~\AA{} and the STO-3G basis set~\cite{Hehre1969}.

Azobenzene provides a molecular application in which the quality of the
2-RDM reconstruction can be examined along a nontrivial excited-state
energy profile~\cite{Wang2024CI,Wang2025cytosine}. We use the 6-31G basis set and an active space of eight
electrons in eight orbitals, [8,8]. Starting from the planar
\emph{trans} ($E$) geometry of $C_{2h}$ symmetry, we perform a rigid scan
of the C--N=N--C dihedral angle $\omega$ from $180^\circ$ to $60^\circ$;
the geometries are not otherwise optimized. Within the active space, we
use the doubly occupied configuration interaction (DOCI) Hamiltonian,
whose exact diagonalization provides the reference ground and excited
states within the seniority-zero sector. The seniority-zero restriction
is used to simplify state preparation on the quantum processor and does
not describe the conical intersection of azobenzene, which requires
seniority-breaking configurations~\cite{pederzoli2011nonadiabatic}.

All quantum circuits are implemented with Qiskit~\cite{Qiskit} and the ffsim library~\cite{ffsim}. The hydrogen-chain calculations use Haar-random orbital rotations as in Ref.~\cite{low2022classical}, while the azobenzene calculations use matchgate shadows implemented as depth-4 nearest-neighbor Givens-rotation circuits on the {\em ibm-phoenix} quantum processor. The semidefinite programs are solved with the Maple Quantum Chemistry Toolbox~\cite{maple_2025,rdmchem_2025} using the boundary-point algorithm of Ref.~\cite{M2011}. Fermionic classical shadows (FCS) are implemented as described in Ref.~\cite{low2022classical}. Unless otherwise stated, FCS and CST use the same total measurement budget, with one shot per randomized unitary for FCS. A value of $\lambda=1$ and a truncation threshold of 0.03 are used throughout this work unless stated otherwise. 

\begin{figure}[htbp]
\centering
\includegraphics[width=\linewidth]{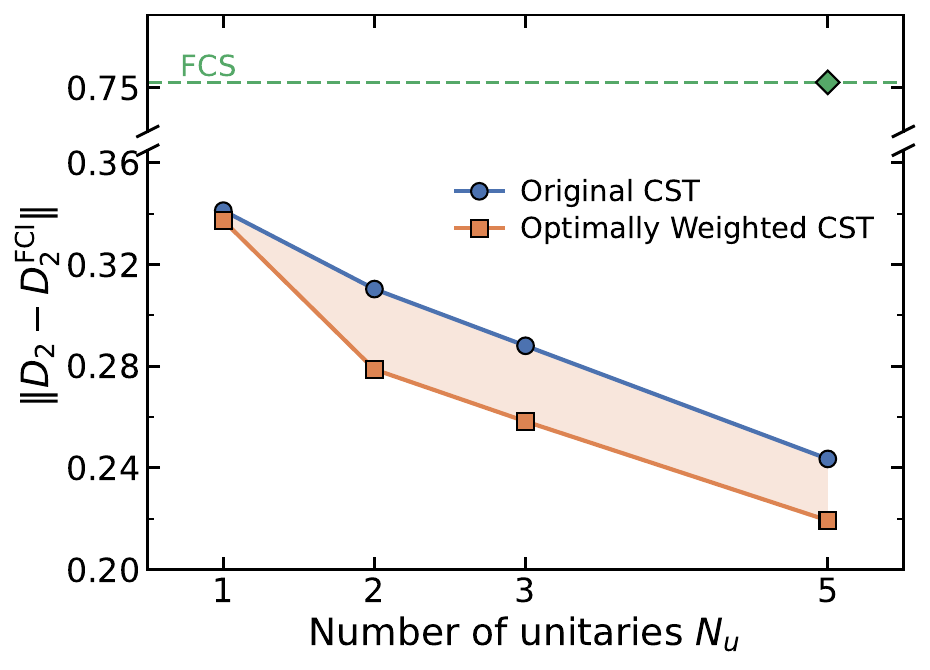}
\caption{2-RDM error $|{}^2D-{}^2D^{\mathrm{FCI}}|$ for H$_6$/STO-3G at $R=1.00$~\AA{} as a function of the number of randomized unitary settings $N_u$ and 10$^4$ shots per setting, comparing the original and optimally weighted constrained shadow tomography (CST) reconstructions and fermionic classical shadows (FCS). Each CST unitary is sampled with $10^4$ shots. FCS is evaluated using 50000 random unitaries and one shot per unitary.}
\label{fig:unitaries}
\end{figure}

Increasing the number of randomized measurement settings systematically improves the reconstruction of the H$_6$ 2-RDM, as shown in Fig.~\ref{fig:unitaries}. Each additional setting provides new measured projections of the 2-RDM. The advantage of optimal weighting becomes apparent when information from multiple randomized settings is combined. In particular, optimally weighted CST with $N_u=2$ yields a smaller 2-RDM error than the original CST reconstruction with $N_u=3$. Thus, weighting the independent measurement directions according to how strongly they are sampled allows comparable or greater accuracy to be obtained from fewer randomized measurement settings.

\begin{figure}[htbp]
\centering
\includegraphics[width=\linewidth]{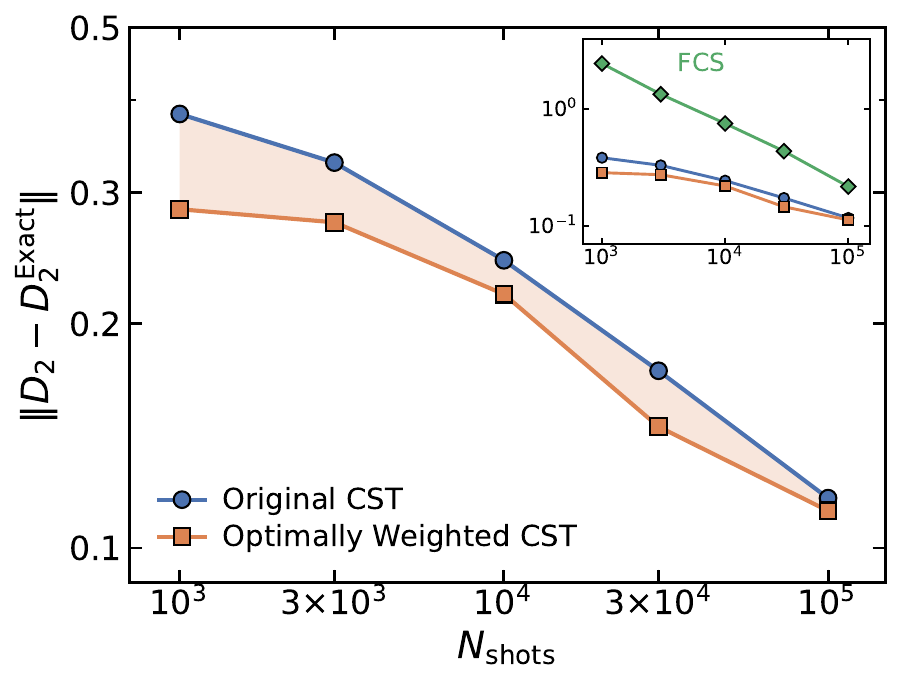}
\caption{Shot-count dependence of the 2-RDM error $|{}^2D-{}^2D^{\mathrm{FCI}}|$ for H$_6$/STO-3G at $R=1.00$~\AA{} with $N_u=5$, comparing the original CST formulation with optimally weighted CST.}
\label{fig:shots}
\end{figure}

The benefit of optimal weighting is most pronounced at finite shot count. Figure~\ref{fig:shots} shows the 2-RDM error for $N_u=5$ as the number of shots per measurement setting is increased from $10^3$ to $10^6$. Optimally weighted CST gives the largest improvement at lower shot counts, where the finite measurement budget produces the greatest differences in how strongly the independent directions of the 2-RDM are sampled. As the number of shots increases, the two reconstructions approach the same limiting accuracy. In the infinite-shot limit, the measured projections are determined exactly, and their relative weighting no longer changes the reconstructed 2-RDM.

\begin{figure}[htbp]
\centering
\includegraphics[width=\linewidth]{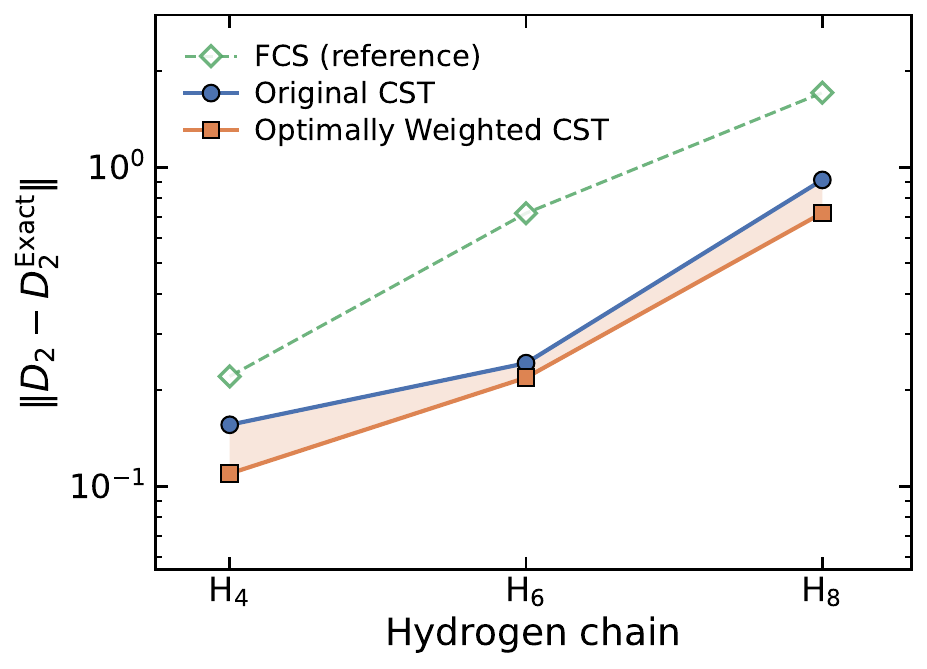}
\caption{2-RDM error $|{}^2D-{}^2D^{\mathrm{exact}}|$ for linear H$_n$ chains at $R=1.00$~\AA{}/STO-3G with $N_u=5$ and $10^4$ shots per setting, comparing the original CST formulation with optimally weighted CST.}
\label{fig:chains}
\end{figure}

The improvement from optimal weighting also persists with increasing system size. For the linear H$_n$ chains in Fig.~\ref{fig:chains}, optimally weighted CST yields a smaller 2-RDM error than the original CST formulation at every chain length examined. Although the absolute Frobenius error increases with system size for both methods, the improvement from optimal weighting is retained across the full series. These results show that the benefit of accounting for how strongly the independent measurement directions are sampled persists as the dimension of the 2-RDM increases.

\begin{figure}[htbp]
    \centering
    \includegraphics[width=\linewidth]{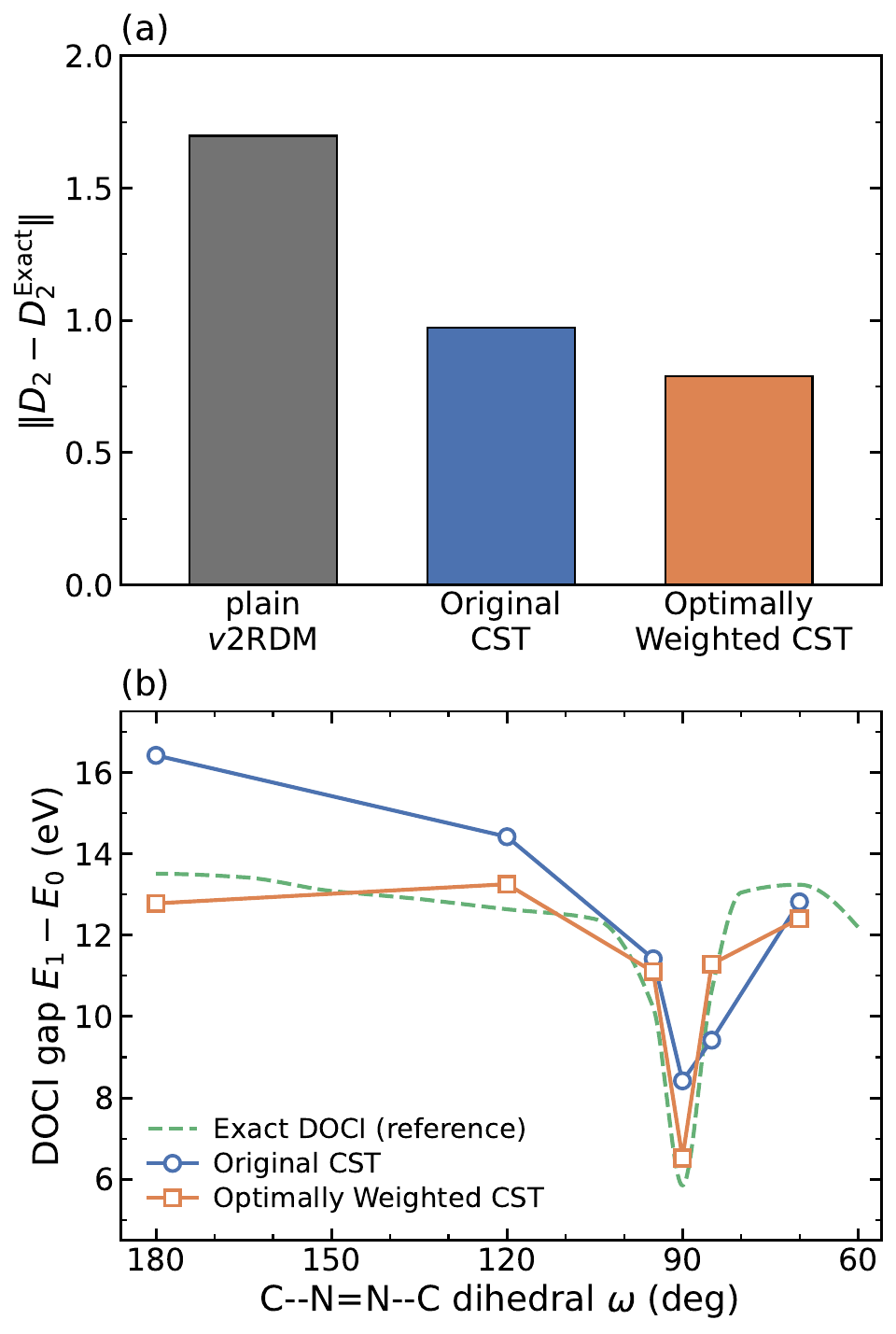}
    \caption{Quantum-processor results for azobenzene in an $[8,8]$
    active space. Each point is obtained using five randomized unitary settings with $10^4$
    shots per setting. (a) Frobenius error of the reconstructed 2-RDM,
    $\|{}^2D-{}^2D^{\mathrm{exact}}\|$, at the planar geometry
    ($\omega=180^\circ$) for variational 2-RDM (v2RDM), the original
    CST formulation, and optimally weighted CST. (b) DOCI excitation
    energy $E_1-E_0$ along a rigid scan of the C--N=N--C dihedral angle
    $\omega$. The dashed curve denotes the exact DOCI result. For excited state, $\lambda=10$ is used to reduce the effect of energy minimization.}
    \label{fig:azo}
\end{figure}

We next apply the method to azobenzene on a quantum processor.
Azobenzene and its derivatives are widely studied photochromic
molecules whose reversible light-driven \emph{cis}--\emph{trans}
isomerization underlies applications ranging from molecular switches
and nanoscale machines to photopharmacology~\cite{Beharry2011,
Garcia2012,Kang2022,Merino2012,Pace2007,Szymanski2013,
pederzoli2011nonadiabatic}. At the planar geometry,
$\omega=180^\circ$, the shadow constraints obtained from the quantum
processor substantially improve the reconstructed 2-RDM relative to the
unconstrained v2RDM result, as shown in Fig.~\ref{fig:azo}(a).
The original CST formulation reduces the error by $43\%$, while
optimally weighted CST reduces it by $54\%$. Thus, the weighting of the
independent measurement directions extracts additional accuracy from
the same quantum measurement data.

The effect of the improved reconstruction is also reflected in the
excited-state energy difference along the torsional coordinate.
As shown in Fig.~\ref{fig:azo}(b), the exact DOCI gap varies strongly
with the C--N=N--C dihedral angle, decreasing from approximately
$13$~eV in the planar region to $5.85$~eV near
$\omega=90^\circ$. Optimally weighted CST follows the reference
profile more closely than the original CST reconstruction over the
scan, demonstrating that the improvement in the 2-RDM carries through
to an energy difference that changes substantially with molecular
geometry. Because the $[8,8]$ DOCI treatment is restricted to the
seniority-zero sector, these calculations are not intended to describe
the physical conical intersection of azobenzene, which requires
seniority-breaking configurations~\cite{pederzoli2011nonadiabatic}.
Rather, the torsional scan provides a chemically relevant test of the
2-RDM reconstruction on quantum hardware.

\section{Discussion and Conclusions}

We have introduced an optimally weighted formulation of constrained shadow tomography for the reconstruction of reduced density matrices. By orthogonalizing the measured constraints, we identify the independent directions of the 2-RDM sampled by the quantum measurements, while the corresponding singular values determine how strongly each direction has been sampled. These weights are incorporated into the $N$-representable semidefinite program through a least-squares objective implemented with a Schur-complement relaxation. In this way, the quantum measurements and the physical constraints on the 2-RDM are combined according to the information actually contained in the measured data. The resulting reconstruction systematically improves the accuracy of the 2-RDM without increasing the quantum measurement budget.

For the hydrogen chains, the improvement from optimal weighting persists with changes in the number of randomized measurement settings, the number of shots, and the system size. For azobenzene, measurements obtained on a quantum processor improve the 2-RDM relative to the variational reconstruction alone, while optimal weighting provides an additional improvement from the same measurement data. The improved reconstruction is also reflected in the excited-state energy difference along the torsional coordinate. These results show that the benefit of optimal weighting extends from controlled model systems to molecular calculations performed with quantum-hardware measurements.

More generally, the approach illustrates how partial quantum measurements can be combined with known physical constraints to reconstruct many-particle quantities more accurately. Whenever measurements probe different independent directions to different extents, the measurement map itself contains information about the appropriate relative importance of those data. Using this information in a physically constrained reconstruction provides a general route for extracting greater accuracy from a fixed quantum measurement budget.

\begin{acknowledgments}
We thank Jamie Garcia for valuable discussions.  D.A.M. gratefully acknowledges support from the U.S. Army Research Office (ARO) under Grant No.~W911NF-26-1-A253 and from IBM through the IBM-UChicago Quantum Collaboration. M.R. gratefully acknowledges the NSF Graduate Research Fellowship Program under grant 2140001.  L.I.P. gratefully acknowledges fellowship support from Jump Trading. Any opinions, findings, and conclusions or recommendations expressed in this material are those of the authors and do not necessarily reflect the views of Jump Trading.  ChatGPT (OpenAI, GPT-5.6 and GPT-6) was used interactively to assist with the refinement of scientific explanations and manuscript editing. The scientific content and conclusions are those of the authors.
\end{acknowledgments}

\section*{Data Availability}

\noindent The data and code supporting the findings of this study are available in Ref.~\cite{WangMazziottiGitHub}.

\bibliography{ref}

\begin{thebibliography}{90}%
\makeatletter
\providecommand \@ifxundefined [1]{%
 \@ifx{#1\undefined}
}%
\providecommand \@ifnum [1]{%
 \ifnum #1\expandafter \@firstoftwo
 \else \expandafter \@secondoftwo
 \fi
}%
\providecommand \@ifx [1]{%
 \ifx #1\expandafter \@firstoftwo
 \else \expandafter \@secondoftwo
 \fi
}%
\providecommand \natexlab [1]{#1}%
\providecommand \enquote  [1]{``#1''}%
\providecommand \bibnamefont  [1]{#1}%
\providecommand \bibfnamefont [1]{#1}%
\providecommand \citenamefont [1]{#1}%
\providecommand \href@noop [0]{\@secondoftwo}%
\providecommand \href [0]{\begingroup \@sanitize@url \@href}%
\providecommand \@href[1]{\@@startlink{#1}\@@href}%
\providecommand \@@href[1]{\endgroup#1\@@endlink}%
\providecommand \@sanitize@url [0]{\catcode `\\12\catcode `\$12\catcode
  `\&12\catcode `\#12\catcode `\^12\catcode `\_12\catcode `\%12\relax}%
\providecommand \@@startlink[1]{}%
\providecommand \@@endlink[0]{}%
\providecommand \url  [0]{\begingroup\@sanitize@url \@url }%
\providecommand \@url [1]{\endgroup\@href {#1}{\urlprefix }}%
\providecommand \urlprefix  [0]{URL }%
\providecommand \Eprint [0]{\href }%
\providecommand \doibase [0]{https://doi.org/}%
\providecommand \selectlanguage [0]{\@gobble}%
\providecommand \bibinfo  [0]{\@secondoftwo}%
\providecommand \bibfield  [0]{\@secondoftwo}%
\providecommand \translation [1]{[#1]}%
\providecommand \BibitemOpen [0]{}%
\providecommand \bibitemStop [0]{}%
\providecommand \bibitemNoStop [0]{.\EOS\space}%
\providecommand \EOS [0]{\spacefactor3000\relax}%
\providecommand \BibitemShut  [1]{\csname bibitem#1\endcsname}%
\let\auto@bib@innerbib\@empty
\bibitem [{\citenamefont {James}\ \emph {et~al.}(2001)\citenamefont {James},
  \citenamefont {Kwiat}, \citenamefont {Munro},\ and\ \citenamefont
  {White}}]{James2001}%
  \BibitemOpen
  \bibfield  {author} {\bibinfo {author} {\bibfnamefont {D.~F.~V.}\
  \bibnamefont {James}}, \bibinfo {author} {\bibfnamefont {P.~G.}\ \bibnamefont
  {Kwiat}}, \bibinfo {author} {\bibfnamefont {W.~J.}\ \bibnamefont {Munro}},\
  and\ \bibinfo {author} {\bibfnamefont {A.~G.}\ \bibnamefont {White}},\
  }\bibfield  {title} {\bibinfo {title} {Measurement of qubits},\ }\href
  {https://doi.org/10.1103/PhysRevA.64.052312} {\bibfield  {journal} {\bibinfo
  {journal} {Phys. Rev. A}\ }\textbf {\bibinfo {volume} {64}},\ \bibinfo
  {pages} {052312} (\bibinfo {year} {2001})}\BibitemShut {NoStop}%
\bibitem [{\citenamefont {Bonet-Monroig}\ \emph {et~al.}(2020)\citenamefont
  {Bonet-Monroig}, \citenamefont {Babbush},\ and\ \citenamefont
  {O’Brien}}]{bonet2020nearly}%
  \BibitemOpen
  \bibfield  {author} {\bibinfo {author} {\bibfnamefont {X.}~\bibnamefont
  {Bonet-Monroig}}, \bibinfo {author} {\bibfnamefont {R.}~\bibnamefont
  {Babbush}},\ and\ \bibinfo {author} {\bibfnamefont {T.~E.}\ \bibnamefont
  {O’Brien}},\ }\bibfield  {title} {\bibinfo {title} {Nearly optimal
  measurement scheduling for partial tomography of quantum states},\ }\href
  {https://doi.org/10.1103/PhysRevX.10.031064} {\bibfield  {journal} {\bibinfo
  {journal} {Phys. Rev. X}\ }\textbf {\bibinfo {volume} {10}},\ \bibinfo
  {pages} {031064} (\bibinfo {year} {2020})}\BibitemShut {NoStop}%
\bibitem [{\citenamefont {Vogel}\ and\ \citenamefont
  {Risken}(1989)}]{Vogel1989}%
  \BibitemOpen
  \bibfield  {author} {\bibinfo {author} {\bibfnamefont {K.}~\bibnamefont
  {Vogel}}\ and\ \bibinfo {author} {\bibfnamefont {H.}~\bibnamefont {Risken}},\
  }\bibfield  {title} {\bibinfo {title} {Determination of quasiprobability
  distributions in terms of probability distributions for the rotated
  quadrature phase},\ }\href {https://doi.org/10.1103/PhysRevA.40.2847}
  {\bibfield  {journal} {\bibinfo  {journal} {Phys. Rev. A}\ }\textbf {\bibinfo
  {volume} {40}},\ \bibinfo {pages} {2847} (\bibinfo {year}
  {1989})}\BibitemShut {NoStop}%
\bibitem [{\citenamefont {Hradil}(1997)}]{Hradil1997}%
  \BibitemOpen
  \bibfield  {author} {\bibinfo {author} {\bibfnamefont {Z.}~\bibnamefont
  {Hradil}},\ }\bibfield  {title} {\bibinfo {title} {Quantum-state
  estimation},\ }\href {https://doi.org/10.1103/PhysRevA.55.R1561} {\bibfield
  {journal} {\bibinfo  {journal} {Phys. Rev. A}\ }\textbf {\bibinfo {volume}
  {55}},\ \bibinfo {pages} {R1561} (\bibinfo {year} {1997})}\BibitemShut
  {NoStop}%
\bibitem [{\citenamefont {Paris}\ and\ \citenamefont
  {\v{R}eh\'{a}\v{c}ek}(2004)}]{Paris2004}%
  \BibitemOpen
  \bibinfo {editor} {\bibfnamefont {M.}~\bibnamefont {Paris}}\ and\ \bibinfo
  {editor} {\bibfnamefont {J.}~\bibnamefont {\v{R}eh\'{a}\v{c}ek}},\ eds.,\
  \href {https://doi.org/10.1007/b98673} {\emph {\bibinfo {title} {Quantum
  State Estimation}}},\ \bibinfo {series} {Lecture Notes in Physics}, Vol.\
  \bibinfo {volume} {649}\ (\bibinfo  {publisher} {Springer},\ \bibinfo
  {address} {Berlin, Heidelberg},\ \bibinfo {year} {2004})\BibitemShut
  {NoStop}%
\bibitem [{\citenamefont {Haah}\ \emph {et~al.}(2017)\citenamefont {Haah},
  \citenamefont {Harrow}, \citenamefont {Ji}, \citenamefont {Wu},\ and\
  \citenamefont {Yu}}]{Haah2017}%
  \BibitemOpen
  \bibfield  {author} {\bibinfo {author} {\bibfnamefont {J.}~\bibnamefont
  {Haah}}, \bibinfo {author} {\bibfnamefont {A.~W.}\ \bibnamefont {Harrow}},
  \bibinfo {author} {\bibfnamefont {Z.}~\bibnamefont {Ji}}, \bibinfo {author}
  {\bibfnamefont {X.}~\bibnamefont {Wu}},\ and\ \bibinfo {author}
  {\bibfnamefont {N.}~\bibnamefont {Yu}},\ }\bibfield  {title} {\bibinfo
  {title} {Sample-optimal tomography of quantum states},\ }\href
  {https://doi.org/10.1109/TIT.2017.2719044} {\bibfield  {journal} {\bibinfo
  {journal} {IEEE Trans. Inf. Theory}\ }\textbf {\bibinfo {volume} {63}},\
  \bibinfo {pages} {5628} (\bibinfo {year} {2017})}\BibitemShut {NoStop}%
\bibitem [{\citenamefont {Aaronson}(2020)}]{Aaronson2020}%
  \BibitemOpen
  \bibfield  {author} {\bibinfo {author} {\bibfnamefont {S.}~\bibnamefont
  {Aaronson}},\ }\bibfield  {title} {\bibinfo {title} {Shadow tomography of
  quantum states},\ }\href {https://doi.org/10.1137/18m120275x} {\bibfield
  {journal} {\bibinfo  {journal} {SIAM J. Comput.}\ }\textbf {\bibinfo {volume}
  {49}},\ \bibinfo {pages} {STOC18\char45 368} (\bibinfo {year}
  {2020})}\BibitemShut {NoStop}%
\bibitem [{\citenamefont {Huang}\ \emph {et~al.}(2020)\citenamefont {Huang},
  \citenamefont {Kueng},\ and\ \citenamefont {Preskill}}]{Huang2020}%
  \BibitemOpen
  \bibfield  {author} {\bibinfo {author} {\bibfnamefont {H.-Y.}\ \bibnamefont
  {Huang}}, \bibinfo {author} {\bibfnamefont {R.}~\bibnamefont {Kueng}},\ and\
  \bibinfo {author} {\bibfnamefont {J.}~\bibnamefont {Preskill}},\ }\bibfield
  {title} {\bibinfo {title} {Predicting many properties of a quantum system
  from very few measurements},\ }\href
  {https://doi.org/10.1038/s41567-020-0932-7} {\bibfield  {journal} {\bibinfo
  {journal} {Nat. Phys.}\ }\textbf {\bibinfo {volume} {16}},\ \bibinfo {pages}
  {1050} (\bibinfo {year} {2020})}\BibitemShut {NoStop}%
\bibitem [{\citenamefont {Zhao}\ \emph {et~al.}(2021)\citenamefont {Zhao},
  \citenamefont {Rubin},\ and\ \citenamefont {Miyake}}]{Zhao2021}%
  \BibitemOpen
  \bibfield  {author} {\bibinfo {author} {\bibfnamefont {A.}~\bibnamefont
  {Zhao}}, \bibinfo {author} {\bibfnamefont {N.~C.}\ \bibnamefont {Rubin}},\
  and\ \bibinfo {author} {\bibfnamefont {A.}~\bibnamefont {Miyake}},\
  }\bibfield  {title} {\bibinfo {title} {Fermionic partial tomography via
  classical shadows},\ }\href {https://doi.org/10.1103/physrevlett.127.110504}
  {\bibfield  {journal} {\bibinfo  {journal} {Phys. Rev. Lett.}\ }\textbf
  {\bibinfo {volume} {127}},\ \bibinfo {pages} {110504} (\bibinfo {year}
  {2021})}\BibitemShut {NoStop}%
\bibitem [{\citenamefont {Elben}\ \emph {et~al.}(2023)\citenamefont {Elben},
  \citenamefont {Flammia}, \citenamefont {Huang}, \citenamefont {Kueng},
  \citenamefont {Preskill}, \citenamefont {Vermersch},\ and\ \citenamefont
  {Zoller}}]{Elben2022}%
  \BibitemOpen
  \bibfield  {author} {\bibinfo {author} {\bibfnamefont {A.}~\bibnamefont
  {Elben}}, \bibinfo {author} {\bibfnamefont {S.~T.}\ \bibnamefont {Flammia}},
  \bibinfo {author} {\bibfnamefont {H.-Y.}\ \bibnamefont {Huang}}, \bibinfo
  {author} {\bibfnamefont {R.}~\bibnamefont {Kueng}}, \bibinfo {author}
  {\bibfnamefont {J.}~\bibnamefont {Preskill}}, \bibinfo {author}
  {\bibfnamefont {B.}~\bibnamefont {Vermersch}},\ and\ \bibinfo {author}
  {\bibfnamefont {P.}~\bibnamefont {Zoller}},\ }\bibfield  {title} {\bibinfo
  {title} {The randomized measurement toolbox},\ }\href
  {https://doi.org/10.1038/s42254-022-00535-2} {\bibfield  {journal} {\bibinfo
  {journal} {Nat. Rev. Phys.}\ }\textbf {\bibinfo {volume} {5}},\ \bibinfo
  {pages} {9} (\bibinfo {year} {2023})}\BibitemShut {NoStop}%
\bibitem [{\citenamefont {Aaronson}\ and\ \citenamefont
  {Rothblum}(2019)}]{Aaronson2019}%
  \BibitemOpen
  \bibfield  {author} {\bibinfo {author} {\bibfnamefont {S.}~\bibnamefont
  {Aaronson}}\ and\ \bibinfo {author} {\bibfnamefont {G.~N.}\ \bibnamefont
  {Rothblum}},\ }\bibfield  {title} {\bibinfo {title} {Gentle measurement of
  quantum states and differential privacy},\ }in\ \href
  {https://doi.org/10.1145/3313276.3316378} {\emph {\bibinfo {booktitle}
  {Proceedings of the 51st Annual ACM SIGACT Symposium on Theory of
  Computing}}}\ (\bibinfo  {publisher} {ACM},\ \bibinfo {address} {New York},\
  \bibinfo {year} {2019})\ pp.\ \bibinfo {pages} {322--333}\BibitemShut
  {NoStop}%
\bibitem [{\citenamefont {Elben}\ \emph {et~al.}(2020)\citenamefont {Elben},
  \citenamefont {Kueng}, \citenamefont {Huang}, \citenamefont {van Bijnen},
  \citenamefont {Kokail}, \citenamefont {Dalmonte}, \citenamefont {Calabrese},
  \citenamefont {Kraus}, \citenamefont {Preskill}, \citenamefont {Zoller},\
  and\ \citenamefont {Vermersch}}]{Elben2020}%
  \BibitemOpen
  \bibfield  {author} {\bibinfo {author} {\bibfnamefont {A.}~\bibnamefont
  {Elben}}, \bibinfo {author} {\bibfnamefont {R.}~\bibnamefont {Kueng}},
  \bibinfo {author} {\bibfnamefont {H.-Y.}\ \bibnamefont {Huang}}, \bibinfo
  {author} {\bibfnamefont {R.}~\bibnamefont {van Bijnen}}, \bibinfo {author}
  {\bibfnamefont {C.}~\bibnamefont {Kokail}}, \bibinfo {author} {\bibfnamefont
  {M.}~\bibnamefont {Dalmonte}}, \bibinfo {author} {\bibfnamefont
  {P.}~\bibnamefont {Calabrese}}, \bibinfo {author} {\bibfnamefont
  {B.}~\bibnamefont {Kraus}}, \bibinfo {author} {\bibfnamefont
  {J.}~\bibnamefont {Preskill}}, \bibinfo {author} {\bibfnamefont
  {P.}~\bibnamefont {Zoller}},\ and\ \bibinfo {author} {\bibfnamefont
  {B.}~\bibnamefont {Vermersch}},\ }\bibfield  {title} {\bibinfo {title}
  {Mixed-state entanglement from local randomized measurements},\ }\href
  {https://doi.org/10.1103/PhysRevLett.125.200501} {\bibfield  {journal}
  {\bibinfo  {journal} {Phys. Rev. Lett.}\ }\textbf {\bibinfo {volume} {125}},\
  \bibinfo {pages} {200501} (\bibinfo {year} {2020})}\BibitemShut {NoStop}%
\bibitem [{\citenamefont {Struchalin}\ \emph {et~al.}(2021)\citenamefont
  {Struchalin}, \citenamefont {Zagorovskii}, \citenamefont {Kovlakov},
  \citenamefont {Straupe},\ and\ \citenamefont {Kulik}}]{Struchalin2021}%
  \BibitemOpen
  \bibfield  {author} {\bibinfo {author} {\bibfnamefont {G.~I.}\ \bibnamefont
  {Struchalin}}, \bibinfo {author} {\bibfnamefont {{\relax Ya}.~A.}\
  \bibnamefont {Zagorovskii}}, \bibinfo {author} {\bibfnamefont {E.~V.}\
  \bibnamefont {Kovlakov}}, \bibinfo {author} {\bibfnamefont {S.~S.}\
  \bibnamefont {Straupe}},\ and\ \bibinfo {author} {\bibfnamefont {S.~P.}\
  \bibnamefont {Kulik}},\ }\bibfield  {title} {\bibinfo {title} {Experimental
  estimation of quantum state properties from classical shadows},\ }\href
  {https://doi.org/10.1103/PRXQuantum.2.010307} {\bibfield  {journal} {\bibinfo
   {journal} {PRX Quantum}\ }\textbf {\bibinfo {volume} {2}},\ \bibinfo {pages}
  {010307} (\bibinfo {year} {2021})}\BibitemShut {NoStop}%
\bibitem [{\citenamefont {B{\u{a}}descu}\ and\ \citenamefont
  {O'Donnell}(2021)}]{Badescu2021}%
  \BibitemOpen
  \bibfield  {author} {\bibinfo {author} {\bibfnamefont {C.}~\bibnamefont
  {B{\u{a}}descu}}\ and\ \bibinfo {author} {\bibfnamefont {R.}~\bibnamefont
  {O'Donnell}},\ }\bibfield  {title} {\bibinfo {title} {Improved quantum data
  analysis},\ }in\ \href {https://doi.org/10.1145/3406325.3451109} {\emph
  {\bibinfo {booktitle} {Proceedings of the 53rd Annual ACM SIGACT Symposium on
  Theory of Computing}}}\ (\bibinfo  {publisher} {ACM},\ \bibinfo {address}
  {New York},\ \bibinfo {year} {2021})\ pp.\ \bibinfo {pages}
  {1398--1411}\BibitemShut {NoStop}%
\bibitem [{\citenamefont {Huang}\ \emph {et~al.}(2021)\citenamefont {Huang},
  \citenamefont {Kueng},\ and\ \citenamefont {Preskill}}]{Huang2021}%
  \BibitemOpen
  \bibfield  {author} {\bibinfo {author} {\bibfnamefont {H.-Y.}\ \bibnamefont
  {Huang}}, \bibinfo {author} {\bibfnamefont {R.}~\bibnamefont {Kueng}},\ and\
  \bibinfo {author} {\bibfnamefont {J.}~\bibnamefont {Preskill}},\ }\bibfield
  {title} {\bibinfo {title} {Efficient estimation of {P}auli observables by
  derandomization},\ }\href {https://doi.org/10.1103/PhysRevLett.127.030503}
  {\bibfield  {journal} {\bibinfo  {journal} {Phys. Rev. Lett.}\ }\textbf
  {\bibinfo {volume} {127}},\ \bibinfo {pages} {030503} (\bibinfo {year}
  {2021})}\BibitemShut {NoStop}%
\bibitem [{\citenamefont {Chen}\ \emph {et~al.}(2021)\citenamefont {Chen},
  \citenamefont {Yu}, \citenamefont {Zeng},\ and\ \citenamefont
  {Flammia}}]{Chen2021}%
  \BibitemOpen
  \bibfield  {author} {\bibinfo {author} {\bibfnamefont {S.}~\bibnamefont
  {Chen}}, \bibinfo {author} {\bibfnamefont {W.}~\bibnamefont {Yu}}, \bibinfo
  {author} {\bibfnamefont {P.}~\bibnamefont {Zeng}},\ and\ \bibinfo {author}
  {\bibfnamefont {S.~T.}\ \bibnamefont {Flammia}},\ }\bibfield  {title}
  {\bibinfo {title} {Robust shadow estimation},\ }\href
  {https://doi.org/10.1103/PRXQuantum.2.030348} {\bibfield  {journal} {\bibinfo
   {journal} {PRX Quantum}\ }\textbf {\bibinfo {volume} {2}},\ \bibinfo {pages}
  {030348} (\bibinfo {year} {2021})}\BibitemShut {NoStop}%
\bibitem [{\citenamefont {Huggins}\ \emph {et~al.}(2022)\citenamefont
  {Huggins}, \citenamefont {O'Gorman}, \citenamefont {Rubin}, \citenamefont
  {Reichman}, \citenamefont {Babbush},\ and\ \citenamefont
  {Lee}}]{Huggins2022}%
  \BibitemOpen
  \bibfield  {author} {\bibinfo {author} {\bibfnamefont {W.~J.}\ \bibnamefont
  {Huggins}}, \bibinfo {author} {\bibfnamefont {B.~A.}\ \bibnamefont
  {O'Gorman}}, \bibinfo {author} {\bibfnamefont {N.~C.}\ \bibnamefont {Rubin}},
  \bibinfo {author} {\bibfnamefont {D.~R.}\ \bibnamefont {Reichman}}, \bibinfo
  {author} {\bibfnamefont {R.}~\bibnamefont {Babbush}},\ and\ \bibinfo {author}
  {\bibfnamefont {J.}~\bibnamefont {Lee}},\ }\bibfield  {title} {\bibinfo
  {title} {Unbiasing fermionic quantum {M}onte {C}arlo with a quantum
  computer},\ }\href {https://doi.org/10.1038/s41586-021-04351-z} {\bibfield
  {journal} {\bibinfo  {journal} {Nature}\ }\textbf {\bibinfo {volume} {603}},\
  \bibinfo {pages} {416} (\bibinfo {year} {2022})}\BibitemShut {NoStop}%
\bibitem [{\citenamefont {Hadfield}\ \emph {et~al.}(2022)\citenamefont
  {Hadfield}, \citenamefont {Bravyi}, \citenamefont {Raymond},\ and\
  \citenamefont {Mezzacapo}}]{Hadfield2022}%
  \BibitemOpen
  \bibfield  {author} {\bibinfo {author} {\bibfnamefont {C.}~\bibnamefont
  {Hadfield}}, \bibinfo {author} {\bibfnamefont {S.}~\bibnamefont {Bravyi}},
  \bibinfo {author} {\bibfnamefont {R.}~\bibnamefont {Raymond}},\ and\ \bibinfo
  {author} {\bibfnamefont {A.}~\bibnamefont {Mezzacapo}},\ }\bibfield  {title}
  {\bibinfo {title} {Measurements of quantum {H}amiltonians with locally-biased
  classical shadows},\ }\href {https://doi.org/10.1007/s00220-022-04343-8}
  {\bibfield  {journal} {\bibinfo  {journal} {Commun. Math. Phys.}\ }\textbf
  {\bibinfo {volume} {391}},\ \bibinfo {pages} {951} (\bibinfo {year}
  {2022})}\BibitemShut {NoStop}%
\bibitem [{\citenamefont {Koh}\ and\ \citenamefont {Grewal}(2022)}]{Koh2022}%
  \BibitemOpen
  \bibfield  {author} {\bibinfo {author} {\bibfnamefont {D.~E.}\ \bibnamefont
  {Koh}}\ and\ \bibinfo {author} {\bibfnamefont {S.}~\bibnamefont {Grewal}},\
  }\bibfield  {title} {\bibinfo {title} {Classical shadows with noise},\ }\href
  {https://doi.org/10.22331/q-2022-08-16-776} {\bibfield  {journal} {\bibinfo
  {journal} {Quantum}\ }\textbf {\bibinfo {volume} {6}},\ \bibinfo {pages}
  {776} (\bibinfo {year} {2022})}\BibitemShut {NoStop}%
\bibitem [{\citenamefont {Bertoni}\ \emph {et~al.}(2024)\citenamefont
  {Bertoni}, \citenamefont {Haferkamp}, \citenamefont {Hinsche}, \citenamefont
  {Ioannou}, \citenamefont {Eisert},\ and\ \citenamefont
  {Pashayan}}]{Bertoni2024}%
  \BibitemOpen
  \bibfield  {author} {\bibinfo {author} {\bibfnamefont {C.}~\bibnamefont
  {Bertoni}}, \bibinfo {author} {\bibfnamefont {J.}~\bibnamefont {Haferkamp}},
  \bibinfo {author} {\bibfnamefont {M.}~\bibnamefont {Hinsche}}, \bibinfo
  {author} {\bibfnamefont {M.}~\bibnamefont {Ioannou}}, \bibinfo {author}
  {\bibfnamefont {J.}~\bibnamefont {Eisert}},\ and\ \bibinfo {author}
  {\bibfnamefont {H.}~\bibnamefont {Pashayan}},\ }\bibfield  {title} {\bibinfo
  {title} {Shallow shadows: {E}xpectation estimation using low-depth random
  {C}lifford circuits},\ }\href
  {https://doi.org/10.1103/PhysRevLett.133.020602} {\bibfield  {journal}
  {\bibinfo  {journal} {Phys. Rev. Lett.}\ }\textbf {\bibinfo {volume} {133}},\
  \bibinfo {pages} {020602} (\bibinfo {year} {2024})}\BibitemShut {NoStop}%
\bibitem [{\citenamefont {Sack}\ \emph {et~al.}(2022)\citenamefont {Sack},
  \citenamefont {Medina}, \citenamefont {Michailidis}, \citenamefont {Kueng},\
  and\ \citenamefont {Serbyn}}]{Sack2022}%
  \BibitemOpen
  \bibfield  {author} {\bibinfo {author} {\bibfnamefont {S.~H.}\ \bibnamefont
  {Sack}}, \bibinfo {author} {\bibfnamefont {R.~A.}\ \bibnamefont {Medina}},
  \bibinfo {author} {\bibfnamefont {A.~A.}\ \bibnamefont {Michailidis}},
  \bibinfo {author} {\bibfnamefont {R.}~\bibnamefont {Kueng}},\ and\ \bibinfo
  {author} {\bibfnamefont {M.}~\bibnamefont {Serbyn}},\ }\bibfield  {title}
  {\bibinfo {title} {Avoiding barren plateaus using classical shadows},\ }\href
  {https://doi.org/10.1103/PRXQuantum.3.020365} {\bibfield  {journal} {\bibinfo
   {journal} {PRX Quantum}\ }\textbf {\bibinfo {volume} {3}},\ \bibinfo {pages}
  {020365} (\bibinfo {year} {2022})}\BibitemShut {NoStop}%
\bibitem [{\citenamefont {Seif}\ \emph {et~al.}(2023)\citenamefont {Seif},
  \citenamefont {Cian}, \citenamefont {Zhou}, \citenamefont {Chen},\ and\
  \citenamefont {Jiang}}]{Seif2023}%
  \BibitemOpen
  \bibfield  {author} {\bibinfo {author} {\bibfnamefont {A.}~\bibnamefont
  {Seif}}, \bibinfo {author} {\bibfnamefont {Z.-P.}\ \bibnamefont {Cian}},
  \bibinfo {author} {\bibfnamefont {S.}~\bibnamefont {Zhou}}, \bibinfo {author}
  {\bibfnamefont {S.}~\bibnamefont {Chen}},\ and\ \bibinfo {author}
  {\bibfnamefont {L.}~\bibnamefont {Jiang}},\ }\bibfield  {title} {\bibinfo
  {title} {Shadow distillation: {Q}uantum error mitigation with classical
  shadows for near-term quantum processors},\ }\href
  {https://doi.org/10.1103/PRXQuantum.4.010303} {\bibfield  {journal} {\bibinfo
   {journal} {PRX Quantum}\ }\textbf {\bibinfo {volume} {4}},\ \bibinfo {pages}
  {010303} (\bibinfo {year} {2023})}\BibitemShut {NoStop}%
\bibitem [{\citenamefont {Hu}\ \emph {et~al.}(2023)\citenamefont {Hu},
  \citenamefont {Choi},\ and\ \citenamefont {You}}]{Hu2023}%
  \BibitemOpen
  \bibfield  {author} {\bibinfo {author} {\bibfnamefont {H.-Y.}\ \bibnamefont
  {Hu}}, \bibinfo {author} {\bibfnamefont {S.}~\bibnamefont {Choi}},\ and\
  \bibinfo {author} {\bibfnamefont {Y.-Z.}\ \bibnamefont {You}},\ }\bibfield
  {title} {\bibinfo {title} {Classical shadow tomography with locally scrambled
  quantum dynamics},\ }\href {https://doi.org/10.1103/PhysRevResearch.5.023027}
  {\bibfield  {journal} {\bibinfo  {journal} {Phys. Rev. Research}\ }\textbf
  {\bibinfo {volume} {5}},\ \bibinfo {pages} {023027} (\bibinfo {year}
  {2023})}\BibitemShut {NoStop}%
\bibitem [{\citenamefont {Levy}\ \emph {et~al.}(2024)\citenamefont {Levy},
  \citenamefont {Luo},\ and\ \citenamefont {Clark}}]{Levy2024}%
  \BibitemOpen
  \bibfield  {author} {\bibinfo {author} {\bibfnamefont {R.}~\bibnamefont
  {Levy}}, \bibinfo {author} {\bibfnamefont {D.}~\bibnamefont {Luo}},\ and\
  \bibinfo {author} {\bibfnamefont {B.~K.}\ \bibnamefont {Clark}},\ }\bibfield
  {title} {\bibinfo {title} {Classical shadows for quantum process tomography
  on near-term quantum computers},\ }\href
  {https://doi.org/10.1103/PhysRevResearch.6.013029} {\bibfield  {journal}
  {\bibinfo  {journal} {Phys. Rev. Research}\ }\textbf {\bibinfo {volume}
  {6}},\ \bibinfo {pages} {013029} (\bibinfo {year} {2024})}\BibitemShut
  {NoStop}%
\bibitem [{\citenamefont {Gidofalvi}\ and\ \citenamefont
  {Mazziotti}(2007)}]{Gidofalvi2007}%
  \BibitemOpen
  \bibfield  {author} {\bibinfo {author} {\bibfnamefont {G.}~\bibnamefont
  {Gidofalvi}}\ and\ \bibinfo {author} {\bibfnamefont {D.~A.}\ \bibnamefont
  {Mazziotti}},\ }\bibfield  {title} {\bibinfo {title} {Molecular properties
  from variational reduced-density-matrix theory with three-particle
  {$N$}-representability conditions},\ }\href
  {https://doi.org/10.1063/1.2423008} {\bibfield  {journal} {\bibinfo
  {journal} {J. Chem. Phys.}\ }\textbf {\bibinfo {volume} {126}},\ \bibinfo
  {pages} {024105} (\bibinfo {year} {2007})}\BibitemShut {NoStop}%
\bibitem [{\citenamefont {Overy}\ \emph {et~al.}(2014)\citenamefont {Overy},
  \citenamefont {Booth}, \citenamefont {Blunt}, \citenamefont {Shepherd},
  \citenamefont {Cleland},\ and\ \citenamefont {Alavi}}]{Overy2014}%
  \BibitemOpen
  \bibfield  {author} {\bibinfo {author} {\bibfnamefont {C.}~\bibnamefont
  {Overy}}, \bibinfo {author} {\bibfnamefont {G.~H.}\ \bibnamefont {Booth}},
  \bibinfo {author} {\bibfnamefont {N.~S.}\ \bibnamefont {Blunt}}, \bibinfo
  {author} {\bibfnamefont {J.~J.}\ \bibnamefont {Shepherd}}, \bibinfo {author}
  {\bibfnamefont {D.}~\bibnamefont {Cleland}},\ and\ \bibinfo {author}
  {\bibfnamefont {A.}~\bibnamefont {Alavi}},\ }\bibfield  {title} {\bibinfo
  {title} {Unbiased reduced density matrices and electronic properties from
  full configuration interaction quantum {M}onte {C}arlo},\ }\href
  {https://doi.org/10.1063/1.4904313} {\bibfield  {journal} {\bibinfo
  {journal} {J. Chem. Phys.}\ }\textbf {\bibinfo {volume} {141}},\ \bibinfo
  {pages} {244117} (\bibinfo {year} {2014})}\BibitemShut {NoStop}%
\bibitem [{\citenamefont {O'Brien}\ \emph {et~al.}(2019)\citenamefont
  {O'Brien}, \citenamefont {Senjean}, \citenamefont {Sagastizabal},
  \citenamefont {Bonet-Monroig}, \citenamefont {Dutkiewicz}, \citenamefont
  {Buda}, \citenamefont {DiCarlo},\ and\ \citenamefont
  {Visscher}}]{OBrien2019}%
  \BibitemOpen
  \bibfield  {author} {\bibinfo {author} {\bibfnamefont {T.~E.}\ \bibnamefont
  {O'Brien}}, \bibinfo {author} {\bibfnamefont {B.}~\bibnamefont {Senjean}},
  \bibinfo {author} {\bibfnamefont {R.}~\bibnamefont {Sagastizabal}}, \bibinfo
  {author} {\bibfnamefont {X.}~\bibnamefont {Bonet-Monroig}}, \bibinfo {author}
  {\bibfnamefont {A.}~\bibnamefont {Dutkiewicz}}, \bibinfo {author}
  {\bibfnamefont {F.}~\bibnamefont {Buda}}, \bibinfo {author} {\bibfnamefont
  {L.}~\bibnamefont {DiCarlo}},\ and\ \bibinfo {author} {\bibfnamefont
  {L.}~\bibnamefont {Visscher}},\ }\bibfield  {title} {\bibinfo {title}
  {Calculating energy derivatives for quantum chemistry on a quantum
  computer},\ }\href {https://doi.org/10.1038/s41534-019-0213-4} {\bibfield
  {journal} {\bibinfo  {journal} {npj Quantum Inf.}\ }\textbf {\bibinfo
  {volume} {5}},\ \bibinfo {pages} {113} (\bibinfo {year} {2019})}\BibitemShut
  {NoStop}%
\bibitem [{\citenamefont {Rubin}\ \emph {et~al.}(2018)\citenamefont {Rubin},
  \citenamefont {Babbush},\ and\ \citenamefont
  {McClean}}]{rubin2018application}%
  \BibitemOpen
  \bibfield  {author} {\bibinfo {author} {\bibfnamefont {N.~C.}\ \bibnamefont
  {Rubin}}, \bibinfo {author} {\bibfnamefont {R.}~\bibnamefont {Babbush}},\
  and\ \bibinfo {author} {\bibfnamefont {J.}~\bibnamefont {McClean}},\
  }\bibfield  {title} {\bibinfo {title} {Application of fermionic marginal
  constraints to hybrid quantum algorithms},\ }\href
  {https://doi.org/10.1088/1367-2630/aab919} {\bibfield  {journal} {\bibinfo
  {journal} {New J. Phys.}\ }\textbf {\bibinfo {volume} {20}},\ \bibinfo
  {pages} {053020} (\bibinfo {year} {2018})}\BibitemShut {NoStop}%
\bibitem [{\citenamefont {Motta}\ \emph {et~al.}(2020)\citenamefont {Motta},
  \citenamefont {Sun}, \citenamefont {Tan}, \citenamefont {O'Rourke},
  \citenamefont {Ye}, \citenamefont {Minnich}, \citenamefont {Brand{\~a}o},\
  and\ \citenamefont {Chan}}]{motta2020qite}%
  \BibitemOpen
  \bibfield  {author} {\bibinfo {author} {\bibfnamefont {M.}~\bibnamefont
  {Motta}}, \bibinfo {author} {\bibfnamefont {C.}~\bibnamefont {Sun}}, \bibinfo
  {author} {\bibfnamefont {A.~T.~K.}\ \bibnamefont {Tan}}, \bibinfo {author}
  {\bibfnamefont {M.~J.}\ \bibnamefont {O'Rourke}}, \bibinfo {author}
  {\bibfnamefont {E.}~\bibnamefont {Ye}}, \bibinfo {author} {\bibfnamefont
  {A.~J.}\ \bibnamefont {Minnich}}, \bibinfo {author} {\bibfnamefont {F.~G.
  S.~L.}\ \bibnamefont {Brand{\~a}o}},\ and\ \bibinfo {author} {\bibfnamefont
  {G.~K.-L.}\ \bibnamefont {Chan}},\ }\bibfield  {title} {\bibinfo {title}
  {Determining eigenstates and thermal states on a quantum computer using
  quantum imaginary time evolution},\ }\href
  {https://doi.org/10.1038/s41567-019-0704-4} {\bibfield  {journal} {\bibinfo
  {journal} {Nat. Phys.}\ }\textbf {\bibinfo {volume} {16}},\ \bibinfo {pages}
  {205} (\bibinfo {year} {2020})}\BibitemShut {NoStop}%
\bibitem [{\citenamefont {Bauer}\ \emph {et~al.}(2016)\citenamefont {Bauer},
  \citenamefont {Wecker}, \citenamefont {Millis}, \citenamefont {Hastings},\
  and\ \citenamefont {Troyer}}]{bauer2016embedding}%
  \BibitemOpen
  \bibfield  {author} {\bibinfo {author} {\bibfnamefont {B.}~\bibnamefont
  {Bauer}}, \bibinfo {author} {\bibfnamefont {D.}~\bibnamefont {Wecker}},
  \bibinfo {author} {\bibfnamefont {A.~J.}\ \bibnamefont {Millis}}, \bibinfo
  {author} {\bibfnamefont {M.~B.}\ \bibnamefont {Hastings}},\ and\ \bibinfo
  {author} {\bibfnamefont {M.}~\bibnamefont {Troyer}},\ }\bibfield  {title}
  {\bibinfo {title} {Hybrid quantum-classical approach to correlated
  materials},\ }\href {https://doi.org/10.1103/PhysRevX.6.031045} {\bibfield
  {journal} {\bibinfo  {journal} {Phys. Rev. X}\ }\textbf {\bibinfo {volume}
  {6}},\ \bibinfo {pages} {031045} (\bibinfo {year} {2016})}\BibitemShut
  {NoStop}%
\bibitem [{\citenamefont {O'Brien}\ \emph {et~al.}(2022)\citenamefont
  {O'Brien}, \citenamefont {Streif}, \citenamefont {Rubin}, \citenamefont
  {Santagati}, \citenamefont {Su}, \citenamefont {Huggins}, \citenamefont
  {Goings}, \citenamefont {Moll}, \citenamefont {Kyoseva}, \citenamefont
  {Degroote}, \citenamefont {Tautermann}, \citenamefont {Lee}, \citenamefont
  {Berry}, \citenamefont {Wiebe},\ and\ \citenamefont
  {Babbush}}]{obrien2022forces}%
  \BibitemOpen
  \bibfield  {author} {\bibinfo {author} {\bibfnamefont {T.~E.}\ \bibnamefont
  {O'Brien}}, \bibinfo {author} {\bibfnamefont {M.}~\bibnamefont {Streif}},
  \bibinfo {author} {\bibfnamefont {N.~C.}\ \bibnamefont {Rubin}}, \bibinfo
  {author} {\bibfnamefont {R.}~\bibnamefont {Santagati}}, \bibinfo {author}
  {\bibfnamefont {Y.}~\bibnamefont {Su}}, \bibinfo {author} {\bibfnamefont
  {W.~J.}\ \bibnamefont {Huggins}}, \bibinfo {author} {\bibfnamefont {J.~J.}\
  \bibnamefont {Goings}}, \bibinfo {author} {\bibfnamefont {N.}~\bibnamefont
  {Moll}}, \bibinfo {author} {\bibfnamefont {E.}~\bibnamefont {Kyoseva}},
  \bibinfo {author} {\bibfnamefont {M.}~\bibnamefont {Degroote}}, \bibinfo
  {author} {\bibfnamefont {C.~S.}\ \bibnamefont {Tautermann}}, \bibinfo
  {author} {\bibfnamefont {J.}~\bibnamefont {Lee}}, \bibinfo {author}
  {\bibfnamefont {D.~W.}\ \bibnamefont {Berry}}, \bibinfo {author}
  {\bibfnamefont {N.}~\bibnamefont {Wiebe}},\ and\ \bibinfo {author}
  {\bibfnamefont {R.}~\bibnamefont {Babbush}},\ }\bibfield  {title} {\bibinfo
  {title} {Efficient quantum computation of molecular forces and other energy
  gradients},\ }\href {https://doi.org/10.1103/PhysRevResearch.4.043210}
  {\bibfield  {journal} {\bibinfo  {journal} {Phys. Rev. Research}\ }\textbf
  {\bibinfo {volume} {4}},\ \bibinfo {pages} {043210} (\bibinfo {year}
  {2022})}\BibitemShut {NoStop}%
\bibitem [{\citenamefont {Smart}\ and\ \citenamefont
  {Mazziotti}(2021)}]{smart2021cqe}%
  \BibitemOpen
  \bibfield  {author} {\bibinfo {author} {\bibfnamefont {S.~E.}\ \bibnamefont
  {Smart}}\ and\ \bibinfo {author} {\bibfnamefont {D.~A.}\ \bibnamefont
  {Mazziotti}},\ }\bibfield  {title} {\bibinfo {title} {Quantum solver of
  contracted eigenvalue equations for scalable molecular simulations on quantum
  computing devices},\ }\href {https://doi.org/10.1103/PhysRevLett.126.070504}
  {\bibfield  {journal} {\bibinfo  {journal} {Phys. Rev. Lett.}\ }\textbf
  {\bibinfo {volume} {126}},\ \bibinfo {pages} {070504} (\bibinfo {year}
  {2021})}\BibitemShut {NoStop}%
\bibitem [{\citenamefont {Wang}\ and\ \citenamefont
  {Mazziotti}(2023)}]{wang2023excited}%
  \BibitemOpen
  \bibfield  {author} {\bibinfo {author} {\bibfnamefont {Y.}~\bibnamefont
  {Wang}}\ and\ \bibinfo {author} {\bibfnamefont {D.~A.}\ \bibnamefont
  {Mazziotti}},\ }\bibfield  {title} {\bibinfo {title} {Electronic excited
  states from a variance-based contracted quantum eigensolver},\ }\href
  {https://doi.org/10.1103/PhysRevA.108.022814} {\bibfield  {journal} {\bibinfo
   {journal} {Phys. Rev. A}\ }\textbf {\bibinfo {volume} {108}},\ \bibinfo
  {pages} {022814} (\bibinfo {year} {2023})}\BibitemShut {NoStop}%
\bibitem [{\citenamefont {Wang}\ and\ \citenamefont
  {Mazziotti}(2025)}]{wang2025rlcqe}%
  \BibitemOpen
  \bibfield  {author} {\bibinfo {author} {\bibfnamefont {Y.}~\bibnamefont
  {Wang}}\ and\ \bibinfo {author} {\bibfnamefont {D.~A.}\ \bibnamefont
  {Mazziotti}},\ }\bibfield  {title} {\bibinfo {title} {Quantum many-body
  simulations from a reinforcement-learned exponential {A}nsatz},\ }\href
  {https://doi.org/10.1103/hrgr-klw3} {\bibfield  {journal} {\bibinfo
  {journal} {Phys. Rev. A}\ }\textbf {\bibinfo {volume} {112}},\ \bibinfo
  {pages} {022403} (\bibinfo {year} {2025})}\BibitemShut {NoStop}%
\bibitem [{\citenamefont {McArdle}\ \emph {et~al.}(2020)\citenamefont
  {McArdle}, \citenamefont {Endo}, \citenamefont {Aspuru-Guzik}, \citenamefont
  {Benjamin},\ and\ \citenamefont {Yuan}}]{McArdle2020}%
  \BibitemOpen
  \bibfield  {author} {\bibinfo {author} {\bibfnamefont {S.}~\bibnamefont
  {McArdle}}, \bibinfo {author} {\bibfnamefont {S.}~\bibnamefont {Endo}},
  \bibinfo {author} {\bibfnamefont {A.}~\bibnamefont {Aspuru-Guzik}}, \bibinfo
  {author} {\bibfnamefont {S.~C.}\ \bibnamefont {Benjamin}},\ and\ \bibinfo
  {author} {\bibfnamefont {X.}~\bibnamefont {Yuan}},\ }\bibfield  {title}
  {\bibinfo {title} {Quantum computational chemistry},\ }\href
  {https://doi.org/10.1103/RevModPhys.92.015003} {\bibfield  {journal}
  {\bibinfo  {journal} {Rev. Mod. Phys.}\ }\textbf {\bibinfo {volume} {92}},\
  \bibinfo {pages} {015003} (\bibinfo {year} {2020})}\BibitemShut {NoStop}%
\bibitem [{\citenamefont {Head-Marsden}\ \emph {et~al.}(2021)\citenamefont
  {Head-Marsden}, \citenamefont {Flick}, \citenamefont {Ciccarino},\ and\
  \citenamefont {Narang}}]{HeadMarsden2021}%
  \BibitemOpen
  \bibfield  {author} {\bibinfo {author} {\bibfnamefont {K.}~\bibnamefont
  {Head-Marsden}}, \bibinfo {author} {\bibfnamefont {J.}~\bibnamefont {Flick}},
  \bibinfo {author} {\bibfnamefont {C.~J.}\ \bibnamefont {Ciccarino}},\ and\
  \bibinfo {author} {\bibfnamefont {P.}~\bibnamefont {Narang}},\ }\bibfield
  {title} {\bibinfo {title} {Quantum information and algorithms for correlated
  quantum matter},\ }\href {https://doi.org/10.1021/acs.chemrev.0c00620}
  {\bibfield  {journal} {\bibinfo  {journal} {Chem. Rev.}\ }\textbf {\bibinfo
  {volume} {121}},\ \bibinfo {pages} {3061} (\bibinfo {year}
  {2021})}\BibitemShut {NoStop}%
\bibitem [{\citenamefont {Jiang}\ \emph {et~al.}(2020)\citenamefont {Jiang},
  \citenamefont {Kalev}, \citenamefont {Mruczkiewicz},\ and\ \citenamefont
  {Neven}}]{Jiang2020}%
  \BibitemOpen
  \bibfield  {author} {\bibinfo {author} {\bibfnamefont {Z.}~\bibnamefont
  {Jiang}}, \bibinfo {author} {\bibfnamefont {A.}~\bibnamefont {Kalev}},
  \bibinfo {author} {\bibfnamefont {W.}~\bibnamefont {Mruczkiewicz}},\ and\
  \bibinfo {author} {\bibfnamefont {H.}~\bibnamefont {Neven}},\ }\bibfield
  {title} {\bibinfo {title} {Optimal fermion-to-qubit mapping via ternary trees
  with applications to reduced quantum states learning},\ }\href
  {https://doi.org/10.22331/q-2020-06-04-276} {\bibfield  {journal} {\bibinfo
  {journal} {Quantum}\ }\textbf {\bibinfo {volume} {4}},\ \bibinfo {pages}
  {276} (\bibinfo {year} {2020})}\BibitemShut {NoStop}%
\bibitem [{\citenamefont {Smart}\ \emph {et~al.}(2022)\citenamefont {Smart},
  \citenamefont {Boyn},\ and\ \citenamefont {Mazziotti}}]{Smart2022}%
  \BibitemOpen
  \bibfield  {author} {\bibinfo {author} {\bibfnamefont {S.~E.}\ \bibnamefont
  {Smart}}, \bibinfo {author} {\bibfnamefont {J.-N.}\ \bibnamefont {Boyn}},\
  and\ \bibinfo {author} {\bibfnamefont {D.~A.}\ \bibnamefont {Mazziotti}},\
  }\bibfield  {title} {\bibinfo {title} {Resolving correlated states of benzyne
  with an error-mitigated contracted quantum eigensolver},\ }\href
  {https://doi.org/10.1103/PhysRevA.105.022405} {\bibfield  {journal} {\bibinfo
   {journal} {Phys. Rev. A}\ }\textbf {\bibinfo {volume} {105}},\ \bibinfo
  {pages} {022405} (\bibinfo {year} {2022})}\BibitemShut {NoStop}%
\bibitem [{\citenamefont {Low}(2022)}]{low2022classical}%
  \BibitemOpen
  \bibfield  {author} {\bibinfo {author} {\bibfnamefont {G.~H.}\ \bibnamefont
  {Low}},\ }\bibfield  {title} {\bibinfo {title} {Classical shadows of fermions
  with particle number symmetry},\ }\bibfield  {journal} {\bibinfo  {journal}
  {arXiv preprint arXiv:2208.08964}\ }\href
  {https://doi.org/10.48550/arXiv.2208.08964} {10.48550/arXiv.2208.08964}
  (\bibinfo {year} {2022})\BibitemShut {NoStop}%
\bibitem [{\citenamefont {Wan}\ \emph {et~al.}(2023)\citenamefont {Wan},
  \citenamefont {Huggins}, \citenamefont {Lee},\ and\ \citenamefont
  {Babbush}}]{Wan2023}%
  \BibitemOpen
  \bibfield  {author} {\bibinfo {author} {\bibfnamefont {K.}~\bibnamefont
  {Wan}}, \bibinfo {author} {\bibfnamefont {W.~J.}\ \bibnamefont {Huggins}},
  \bibinfo {author} {\bibfnamefont {J.}~\bibnamefont {Lee}},\ and\ \bibinfo
  {author} {\bibfnamefont {R.}~\bibnamefont {Babbush}},\ }\bibfield  {title}
  {\bibinfo {title} {Matchgate shadows for fermionic quantum simulation},\
  }\href {https://doi.org/10.1007/s00220-023-04844-0} {\bibfield  {journal}
  {\bibinfo  {journal} {Commun. Math. Phys.}\ }\textbf {\bibinfo {volume}
  {404}},\ \bibinfo {pages} {629} (\bibinfo {year} {2023})}\BibitemShut
  {NoStop}%
\bibitem [{\citenamefont {Huang}\ \emph {et~al.}(2024)\citenamefont {Huang},
  \citenamefont {Chen}, \citenamefont {Gupt}, \citenamefont {Suchara},
  \citenamefont {Tran}, \citenamefont {McArdle},\ and\ \citenamefont
  {Galli}}]{Huang2024}%
  \BibitemOpen
  \bibfield  {author} {\bibinfo {author} {\bibfnamefont {B.}~\bibnamefont
  {Huang}}, \bibinfo {author} {\bibfnamefont {Y.-T.}\ \bibnamefont {Chen}},
  \bibinfo {author} {\bibfnamefont {B.}~\bibnamefont {Gupt}}, \bibinfo {author}
  {\bibfnamefont {M.}~\bibnamefont {Suchara}}, \bibinfo {author} {\bibfnamefont
  {A.}~\bibnamefont {Tran}}, \bibinfo {author} {\bibfnamefont {S.}~\bibnamefont
  {McArdle}},\ and\ \bibinfo {author} {\bibfnamefont {G.}~\bibnamefont
  {Galli}},\ }\bibfield  {title} {\bibinfo {title} {Evaluating a
  quantum-classical quantum {M}onte {C}arlo algorithm with matchgate shadows},\
  }\href {https://doi.org/10.1103/PhysRevResearch.6.043063} {\bibfield
  {journal} {\bibinfo  {journal} {Phys. Rev. Research}\ }\textbf {\bibinfo
  {volume} {6}},\ \bibinfo {pages} {043063} (\bibinfo {year}
  {2024})}\BibitemShut {NoStop}%
\bibitem [{\citenamefont {Zhao}\ and\ \citenamefont
  {Miyake}(2024)}]{ZhaoMiyake2024}%
  \BibitemOpen
  \bibfield  {author} {\bibinfo {author} {\bibfnamefont {A.}~\bibnamefont
  {Zhao}}\ and\ \bibinfo {author} {\bibfnamefont {A.}~\bibnamefont {Miyake}},\
  }\bibfield  {title} {\bibinfo {title} {Group-theoretic error mitigation
  enabled by classical shadows and symmetries},\ }\href
  {https://doi.org/10.1038/s41534-024-00854-5} {\bibfield  {journal} {\bibinfo
  {journal} {npj Quantum Inf.}\ }\textbf {\bibinfo {volume} {10}},\ \bibinfo
  {pages} {57} (\bibinfo {year} {2024})}\BibitemShut {NoStop}%
\bibitem [{\citenamefont {King}\ \emph {et~al.}(2025)\citenamefont {King},
  \citenamefont {Gosset}, \citenamefont {Kothari},\ and\ \citenamefont
  {Babbush}}]{King2025}%
  \BibitemOpen
  \bibfield  {author} {\bibinfo {author} {\bibfnamefont {R.}~\bibnamefont
  {King}}, \bibinfo {author} {\bibfnamefont {D.}~\bibnamefont {Gosset}},
  \bibinfo {author} {\bibfnamefont {R.}~\bibnamefont {Kothari}},\ and\ \bibinfo
  {author} {\bibfnamefont {R.}~\bibnamefont {Babbush}},\ }\bibfield  {title}
  {\bibinfo {title} {Triply efficient shadow tomography},\ }\href
  {https://doi.org/10.1103/PRXQuantum.6.010336} {\bibfield  {journal} {\bibinfo
   {journal} {PRX Quantum}\ }\textbf {\bibinfo {volume} {6}},\ \bibinfo {pages}
  {010336} (\bibinfo {year} {2025})}\BibitemShut {NoStop}%
\bibitem [{\citenamefont {Wang}\ \emph
  {et~al.}(2025{\natexlab{a}})\citenamefont {Wang}, \citenamefont {Avdic},\
  and\ \citenamefont {Mazziotti}}]{wang2025shadow}%
  \BibitemOpen
  \bibfield  {author} {\bibinfo {author} {\bibfnamefont {Y.}~\bibnamefont
  {Wang}}, \bibinfo {author} {\bibfnamefont {I.}~\bibnamefont {Avdic}},\ and\
  \bibinfo {author} {\bibfnamefont {D.~A.}\ \bibnamefont {Mazziotti}},\
  }\bibfield  {title} {\bibinfo {title} {Shadow ansatz for the many-fermion
  wave function in scalable molecular simulations on quantum computing
  devices},\ }\href {https://doi.org/10.1103/1nml-xjqw} {\bibfield  {journal}
  {\bibinfo  {journal} {Phys. Rev. A}\ }\textbf {\bibinfo {volume} {112}},\
  \bibinfo {pages} {022432} (\bibinfo {year} {2025}{\natexlab{a}})}\BibitemShut
  {NoStop}%
\bibitem [{\citenamefont {Coleman}(1963)}]{Coleman1963}%
  \BibitemOpen
  \bibfield  {author} {\bibinfo {author} {\bibfnamefont {A.~J.}\ \bibnamefont
  {Coleman}},\ }\bibfield  {title} {\bibinfo {title} {Structure of fermion
  density matrices},\ }\href {https://doi.org/10.1103/RevModPhys.35.668}
  {\bibfield  {journal} {\bibinfo  {journal} {Rev. Mod. Phys.}\ }\textbf
  {\bibinfo {volume} {35}},\ \bibinfo {pages} {668} (\bibinfo {year}
  {1963})}\BibitemShut {NoStop}%
\bibitem [{\citenamefont {Kummer}(1967)}]{Kummer1967}%
  \BibitemOpen
  \bibfield  {author} {\bibinfo {author} {\bibfnamefont {H.}~\bibnamefont
  {Kummer}},\ }\bibfield  {title} {\bibinfo {title} {{$N$}-representability
  problem for reduced density matrices},\ }\href
  {https://doi.org/10.1063/1.1705122} {\bibfield  {journal} {\bibinfo
  {journal} {J. Math. Phys.}\ }\textbf {\bibinfo {volume} {8}},\ \bibinfo
  {pages} {2063} (\bibinfo {year} {1967})}\BibitemShut {NoStop}%
\bibitem [{\citenamefont {Erdahl}(1978)}]{Erdahl1978}%
  \BibitemOpen
  \bibfield  {author} {\bibinfo {author} {\bibfnamefont {R.~M.}\ \bibnamefont
  {Erdahl}},\ }\bibfield  {title} {\bibinfo {title} {Representability},\ }\href
  {https://doi.org/10.1002/qua.560130603} {\bibfield  {journal} {\bibinfo
  {journal} {Int. J. Quantum Chem.}\ }\textbf {\bibinfo {volume} {13}},\
  \bibinfo {pages} {697} (\bibinfo {year} {1978})}\BibitemShut {NoStop}%
\bibitem [{\citenamefont {Coleman}\ and\ \citenamefont
  {Yukalov}(2000)}]{ColemanYukalov2000}%
  \BibitemOpen
  \bibfield  {author} {\bibinfo {author} {\bibfnamefont {A.~J.}\ \bibnamefont
  {Coleman}}\ and\ \bibinfo {author} {\bibfnamefont {V.~I.}\ \bibnamefont
  {Yukalov}},\ }\href {https://doi.org/10.1007/978-3-642-58304-9} {\emph
  {\bibinfo {title} {Reduced Density Matrices: Coulson's Challenge}}},\
  \bibinfo {series} {Lecture Notes in Chemistry}, Vol.~\bibinfo {volume} {72}\
  (\bibinfo  {publisher} {Springer},\ \bibinfo {address} {Berlin, Heidelberg},\
  \bibinfo {year} {2000})\BibitemShut {NoStop}%
\bibitem [{\citenamefont {Avdic}\ and\ \citenamefont
  {Mazziotti}(2024{\natexlab{a}})}]{Avdic2024}%
  \BibitemOpen
  \bibfield  {author} {\bibinfo {author} {\bibfnamefont {I.}~\bibnamefont
  {Avdic}}\ and\ \bibinfo {author} {\bibfnamefont {D.~A.}\ \bibnamefont
  {Mazziotti}},\ }\bibfield  {title} {\bibinfo {title} {Fewer measurements from
  shadow tomography with {$N$}-representability conditions},\ }\href
  {https://doi.org/10.1103/physrevlett.132.220802} {\bibfield  {journal}
  {\bibinfo  {journal} {Phys. Rev. Lett.}\ }\textbf {\bibinfo {volume} {132}},\
  \bibinfo {pages} {220802} (\bibinfo {year} {2024}{\natexlab{a}})}\BibitemShut
  {NoStop}%
\bibitem [{\citenamefont {Avdic}\ and\ \citenamefont
  {Mazziotti}(2024{\natexlab{b}})}]{Avdic2024excited}%
  \BibitemOpen
  \bibfield  {author} {\bibinfo {author} {\bibfnamefont {I.}~\bibnamefont
  {Avdic}}\ and\ \bibinfo {author} {\bibfnamefont {D.~A.}\ \bibnamefont
  {Mazziotti}},\ }\bibfield  {title} {\bibinfo {title} {Enhanced shadow
  tomography of molecular excited states via the enforcement of
  {$N$}-representability conditions by semidefinite programming},\ }\href
  {https://doi.org/10.1103/PhysRevA.110.052407} {\bibfield  {journal} {\bibinfo
   {journal} {Phys. Rev. A}\ }\textbf {\bibinfo {volume} {110}},\ \bibinfo
  {pages} {052407} (\bibinfo {year} {2024}{\natexlab{b}})}\BibitemShut
  {NoStop}%
\bibitem [{\citenamefont {Avdic}\ \emph {et~al.}(2026)\citenamefont {Avdic},
  \citenamefont {Wang}, \citenamefont {Rose}, \citenamefont {Payne~Torres},
  \citenamefont {Schouten}, \citenamefont {Sung},\ and\ \citenamefont
  {Mazziotti}}]{avdic2026constrained}%
  \BibitemOpen
  \bibfield  {author} {\bibinfo {author} {\bibfnamefont {I.}~\bibnamefont
  {Avdic}}, \bibinfo {author} {\bibfnamefont {Y.}~\bibnamefont {Wang}},
  \bibinfo {author} {\bibfnamefont {M.}~\bibnamefont {Rose}}, \bibinfo {author}
  {\bibfnamefont {L.~I.}\ \bibnamefont {Payne~Torres}}, \bibinfo {author}
  {\bibfnamefont {A.~O.}\ \bibnamefont {Schouten}}, \bibinfo {author}
  {\bibfnamefont {K.~J.}\ \bibnamefont {Sung}},\ and\ \bibinfo {author}
  {\bibfnamefont {D.~A.}\ \bibnamefont {Mazziotti}},\ }\bibfield  {title}
  {\bibinfo {title} {Constrained shadow tomography for molecular simulation on
  quantum devices},\ }\href {https://doi.org/10.1039/D5SC09162D} {\bibfield
  {journal} {\bibinfo  {journal} {Chem. Sci.}\ }\textbf {\bibinfo {volume}
  {17}},\ \bibinfo {pages} {9633} (\bibinfo {year} {2026})}\BibitemShut
  {NoStop}%
\bibitem [{\citenamefont {Wang}\ \emph {et~al.}(2026)\citenamefont {Wang},
  \citenamefont {Avdic}, \citenamefont {Rose}, \citenamefont {Payne~Torres},
  \citenamefont {Schouten}, \citenamefont {Sung},\ and\ \citenamefont
  {Mazziotti}}]{wang2026correlated}%
  \BibitemOpen
  \bibfield  {author} {\bibinfo {author} {\bibfnamefont {Y.}~\bibnamefont
  {Wang}}, \bibinfo {author} {\bibfnamefont {I.}~\bibnamefont {Avdic}},
  \bibinfo {author} {\bibfnamefont {M.}~\bibnamefont {Rose}}, \bibinfo {author}
  {\bibfnamefont {L.~I.}\ \bibnamefont {Payne~Torres}}, \bibinfo {author}
  {\bibfnamefont {A.~O.}\ \bibnamefont {Schouten}}, \bibinfo {author}
  {\bibfnamefont {K.~J.}\ \bibnamefont {Sung}},\ and\ \bibinfo {author}
  {\bibfnamefont {D.~A.}\ \bibnamefont {Mazziotti}},\ }\bibfield  {title}
  {\bibinfo {title} {Correlated purification for restoring
  {$N$}-representability in quantum simulation},\ }\href
  {https://doi.org/10.1103/hwhj-b9cq} {\bibfield  {journal} {\bibinfo
  {journal} {Phys. Rev. A}\ }\textbf {\bibinfo {volume} {113}},\ \bibinfo
  {pages} {032424} (\bibinfo {year} {2026})}\BibitemShut {NoStop}%
\bibitem [{\citenamefont {Mazziotti}\ and\ \citenamefont
  {Erdahl}(2001)}]{MazziottiErdahl2001}%
  \BibitemOpen
  \bibfield  {author} {\bibinfo {author} {\bibfnamefont {D.~A.}\ \bibnamefont
  {Mazziotti}}\ and\ \bibinfo {author} {\bibfnamefont {R.~M.}\ \bibnamefont
  {Erdahl}},\ }\bibfield  {title} {\bibinfo {title} {Uncertainty relations and
  reduced density matrices: {M}apping many-body quantum mechanics onto four
  particles},\ }\href {https://doi.org/10.1103/PhysRevA.63.042113} {\bibfield
  {journal} {\bibinfo  {journal} {Phys. Rev. A}\ }\textbf {\bibinfo {volume}
  {63}},\ \bibinfo {pages} {042113} (\bibinfo {year} {2001})}\BibitemShut
  {NoStop}%
\bibitem [{\citenamefont {Nakata}\ \emph {et~al.}(2001)\citenamefont {Nakata},
  \citenamefont {Nakatsuji}, \citenamefont {Ehara}, \citenamefont {Fukuda},
  \citenamefont {Nakata},\ and\ \citenamefont {Fujisawa}}]{Nakata2001}%
  \BibitemOpen
  \bibfield  {author} {\bibinfo {author} {\bibfnamefont {M.}~\bibnamefont
  {Nakata}}, \bibinfo {author} {\bibfnamefont {H.}~\bibnamefont {Nakatsuji}},
  \bibinfo {author} {\bibfnamefont {M.}~\bibnamefont {Ehara}}, \bibinfo
  {author} {\bibfnamefont {M.}~\bibnamefont {Fukuda}}, \bibinfo {author}
  {\bibfnamefont {K.}~\bibnamefont {Nakata}},\ and\ \bibinfo {author}
  {\bibfnamefont {K.}~\bibnamefont {Fujisawa}},\ }\bibfield  {title} {\bibinfo
  {title} {Variational calculations of fermion second-order reduced density
  matrices by semidefinite programming algorithm},\ }\href
  {https://doi.org/10.1063/1.1360199} {\bibfield  {journal} {\bibinfo
  {journal} {J. Chem. Phys.}\ }\textbf {\bibinfo {volume} {114}},\ \bibinfo
  {pages} {8282} (\bibinfo {year} {2001})}\BibitemShut {NoStop}%
\bibitem [{\citenamefont {Mazziotti}(2002)}]{Mazziotti2002}%
  \BibitemOpen
  \bibfield  {author} {\bibinfo {author} {\bibfnamefont {D.~A.}\ \bibnamefont
  {Mazziotti}},\ }\bibfield  {title} {\bibinfo {title} {Variational
  minimization of atomic and molecular ground-state energies via the
  two-particle reduced density matrix},\ }\href
  {https://doi.org/10.1103/PhysRevA.65.062511} {\bibfield  {journal} {\bibinfo
  {journal} {Phys. Rev. A}\ }\textbf {\bibinfo {volume} {65}},\ \bibinfo
  {pages} {062511} (\bibinfo {year} {2002})}\BibitemShut {NoStop}%
\bibitem [{\citenamefont {Mazziotti}(2004)}]{M2004}%
  \BibitemOpen
  \bibfield  {author} {\bibinfo {author} {\bibfnamefont {D.~A.}\ \bibnamefont
  {Mazziotti}},\ }\bibfield  {title} {\bibinfo {title} {Realization of quantum
  chemistry without wave functions through first-order semidefinite
  programming},\ }\href {https://doi.org/10.1103/PhysRevLett.93.213001}
  {\bibfield  {journal} {\bibinfo  {journal} {Phys. Rev. Lett.}\ }\textbf
  {\bibinfo {volume} {93}},\ \bibinfo {pages} {213001} (\bibinfo {year}
  {2004})}\BibitemShut {NoStop}%
\bibitem [{\citenamefont {Zhao}\ \emph {et~al.}(2004)\citenamefont {Zhao},
  \citenamefont {Braams}, \citenamefont {Fukuda}, \citenamefont {Overton},\
  and\ \citenamefont {Percus}}]{Zhao2004}%
  \BibitemOpen
  \bibfield  {author} {\bibinfo {author} {\bibfnamefont {Z.}~\bibnamefont
  {Zhao}}, \bibinfo {author} {\bibfnamefont {B.~J.}\ \bibnamefont {Braams}},
  \bibinfo {author} {\bibfnamefont {M.}~\bibnamefont {Fukuda}}, \bibinfo
  {author} {\bibfnamefont {M.~L.}\ \bibnamefont {Overton}},\ and\ \bibinfo
  {author} {\bibfnamefont {J.~K.}\ \bibnamefont {Percus}},\ }\bibfield  {title}
  {\bibinfo {title} {The reduced density matrix method for electronic structure
  calculations and the role of three-index representability conditions},\
  }\href {https://doi.org/10.1063/1.1636721} {\bibfield  {journal} {\bibinfo
  {journal} {J. Chem. Phys.}\ }\textbf {\bibinfo {volume} {120}},\ \bibinfo
  {pages} {2095} (\bibinfo {year} {2004})}\BibitemShut {NoStop}%
\bibitem [{\citenamefont {Gidofalvi}\ and\ \citenamefont
  {Mazziotti}(2005)}]{Gidofalvi2005}%
  \BibitemOpen
  \bibfield  {author} {\bibinfo {author} {\bibfnamefont {G.}~\bibnamefont
  {Gidofalvi}}\ and\ \bibinfo {author} {\bibfnamefont {D.~A.}\ \bibnamefont
  {Mazziotti}},\ }\bibfield  {title} {\bibinfo {title} {Spin and symmetry
  adaptation of the variational two-electron reduced-density-matrix method},\
  }\href {https://doi.org/10.1103/PhysRevA.72.052505} {\bibfield  {journal}
  {\bibinfo  {journal} {Phys. Rev. A}\ }\textbf {\bibinfo {volume} {72}},\
  \bibinfo {pages} {052505} (\bibinfo {year} {2005})}\BibitemShut {NoStop}%
\bibitem [{\citenamefont {Canc{\`e}s}\ \emph {et~al.}(2006)\citenamefont
  {Canc{\`e}s}, \citenamefont {Stoltz},\ and\ \citenamefont
  {Lewin}}]{Cances2006}%
  \BibitemOpen
  \bibfield  {author} {\bibinfo {author} {\bibfnamefont {E.}~\bibnamefont
  {Canc{\`e}s}}, \bibinfo {author} {\bibfnamefont {G.}~\bibnamefont {Stoltz}},\
  and\ \bibinfo {author} {\bibfnamefont {M.}~\bibnamefont {Lewin}},\ }\bibfield
   {title} {\bibinfo {title} {The electronic ground-state energy problem: {A}
  new reduced density matrix approach},\ }\href
  {https://doi.org/10.1063/1.2222358} {\bibfield  {journal} {\bibinfo
  {journal} {J. Chem. Phys.}\ }\textbf {\bibinfo {volume} {125}},\ \bibinfo
  {pages} {064101} (\bibinfo {year} {2006})}\BibitemShut {NoStop}%
\bibitem [{\citenamefont {Gidofalvi}\ and\ \citenamefont
  {Mazziotti}(2008)}]{gidofalvi2008activespace}%
  \BibitemOpen
  \bibfield  {author} {\bibinfo {author} {\bibfnamefont {G.}~\bibnamefont
  {Gidofalvi}}\ and\ \bibinfo {author} {\bibfnamefont {D.~A.}\ \bibnamefont
  {Mazziotti}},\ }\bibfield  {title} {\bibinfo {title} {Active-space
  two-electron reduced-density-matrix method: Complete active-space
  calculations without diagonalization of the {$N$}-electron {H}amiltonian},\
  }\href {https://doi.org/10.1063/1.2983652} {\bibfield  {journal} {\bibinfo
  {journal} {J. Chem. Phys.}\ }\textbf {\bibinfo {volume} {129}},\ \bibinfo
  {pages} {134108} (\bibinfo {year} {2008})}\BibitemShut {NoStop}%
\bibitem [{\citenamefont {Shenvi}\ and\ \citenamefont
  {Izmaylov}(2010)}]{Shenvi2010}%
  \BibitemOpen
  \bibfield  {author} {\bibinfo {author} {\bibfnamefont {N.}~\bibnamefont
  {Shenvi}}\ and\ \bibinfo {author} {\bibfnamefont {A.~F.}\ \bibnamefont
  {Izmaylov}},\ }\bibfield  {title} {\bibinfo {title} {Active-space
  {$N$}-representability constraints for variational two-particle reduced
  density matrix calculations},\ }\href
  {https://doi.org/10.1103/PhysRevLett.105.213003} {\bibfield  {journal}
  {\bibinfo  {journal} {Phys. Rev. Lett.}\ }\textbf {\bibinfo {volume} {105}},\
  \bibinfo {pages} {213003} (\bibinfo {year} {2010})}\BibitemShut {NoStop}%
\bibitem [{\citenamefont {Mazziotti}(2016)}]{Mazziotti2016}%
  \BibitemOpen
  \bibfield  {author} {\bibinfo {author} {\bibfnamefont {D.~A.}\ \bibnamefont
  {Mazziotti}},\ }\bibfield  {title} {\bibinfo {title} {Enhanced constraints
  for accurate lower bounds on many-electron quantum energies from variational
  two-electron reduced density matrix theory},\ }\href
  {https://doi.org/10.1103/PhysRevLett.117.153001} {\bibfield  {journal}
  {\bibinfo  {journal} {Phys. Rev. Lett.}\ }\textbf {\bibinfo {volume} {117}},\
  \bibinfo {pages} {153001} (\bibinfo {year} {2016})}\BibitemShut {NoStop}%
\bibitem [{\citenamefont {Alcoba}\ \emph {et~al.}(2018)\citenamefont {Alcoba},
  \citenamefont {Torre}, \citenamefont {Lain}, \citenamefont {Massaccesi},
  \citenamefont {O{\~n}a}, \citenamefont {Honor{\'e}}, \citenamefont
  {Poelmans}, \citenamefont {Van~Neck}, \citenamefont {Bultinck},\ and\
  \citenamefont {De~Baerdemacker}}]{Alcoba2018}%
  \BibitemOpen
  \bibfield  {author} {\bibinfo {author} {\bibfnamefont {D.~R.}\ \bibnamefont
  {Alcoba}}, \bibinfo {author} {\bibfnamefont {A.}~\bibnamefont {Torre}},
  \bibinfo {author} {\bibfnamefont {L.}~\bibnamefont {Lain}}, \bibinfo {author}
  {\bibfnamefont {G.~E.}\ \bibnamefont {Massaccesi}}, \bibinfo {author}
  {\bibfnamefont {O.~B.}\ \bibnamefont {O{\~n}a}}, \bibinfo {author}
  {\bibfnamefont {E.~M.}\ \bibnamefont {Honor{\'e}}}, \bibinfo {author}
  {\bibfnamefont {W.}~\bibnamefont {Poelmans}}, \bibinfo {author}
  {\bibfnamefont {D.}~\bibnamefont {Van~Neck}}, \bibinfo {author}
  {\bibfnamefont {P.}~\bibnamefont {Bultinck}},\ and\ \bibinfo {author}
  {\bibfnamefont {S.}~\bibnamefont {De~Baerdemacker}},\ }\bibfield  {title}
  {\bibinfo {title} {Direct variational determination of the two-electron
  reduced density matrix for doubly occupied-configuration-interaction wave
  functions: {T}he influence of three-index {$N$}-representability
  conditions},\ }\href {https://doi.org/10.1063/1.5008811} {\bibfield
  {journal} {\bibinfo  {journal} {J. Chem. Phys.}\ }\textbf {\bibinfo {volume}
  {148}},\ \bibinfo {pages} {024105} (\bibinfo {year} {2018})}\BibitemShut
  {NoStop}%
\bibitem [{\citenamefont {Mazziotti}(2007)}]{M2007}%
  \BibitemOpen
  \bibinfo {editor} {\bibfnamefont {D.~A.}\ \bibnamefont {Mazziotti}},\ ed.,\
  \href {https://doi.org/10.1002/0470106603} {\emph {\bibinfo {title}
  {Reduced-Density-Matrix Mechanics: With Application to Many-Electron Atoms
  and Molecules}}},\ \bibinfo {series} {Advances in Chemical Physics}, Vol.\
  \bibinfo {volume} {134}\ (\bibinfo  {publisher} {Wiley},\ \bibinfo {address}
  {Hoboken, NJ},\ \bibinfo {year} {2007})\BibitemShut {NoStop}%
\bibitem [{\citenamefont {Mazziotti}(2012{\natexlab{a}})}]{M2012}%
  \BibitemOpen
  \bibfield  {author} {\bibinfo {author} {\bibfnamefont {D.~A.}\ \bibnamefont
  {Mazziotti}},\ }\bibfield  {title} {\bibinfo {title} {Structure of fermionic
  density matrices: Complete ${N}$-representability conditions},\ }\href
  {https://doi.org/10.1103/PhysRevLett.108.263002} {\bibfield  {journal}
  {\bibinfo  {journal} {Phys. Rev. Lett.}\ }\textbf {\bibinfo {volume} {108}},\
  \bibinfo {pages} {263002} (\bibinfo {year} {2012}{\natexlab{a}})}\BibitemShut
  {NoStop}%
\bibitem [{\citenamefont {Mazziotti}(2012{\natexlab{b}})}]{M2012b}%
  \BibitemOpen
  \bibfield  {author} {\bibinfo {author} {\bibfnamefont {D.~A.}\ \bibnamefont
  {Mazziotti}},\ }\bibfield  {title} {\bibinfo {title} {Two-electron reduced
  density matrix as the basic variable in many-electron quantum chemistry and
  physics},\ }\href {https://doi.org/10.1021/cr2000493} {\bibfield  {journal}
  {\bibinfo  {journal} {Chem. Rev.}\ }\textbf {\bibinfo {volume} {112}},\
  \bibinfo {pages} {244} (\bibinfo {year} {2012}{\natexlab{b}})}\BibitemShut
  {NoStop}%
\bibitem [{\citenamefont {Garrod}\ and\ \citenamefont
  {Percus}(1964)}]{Garrod1964}%
  \BibitemOpen
  \bibfield  {author} {\bibinfo {author} {\bibfnamefont {C.}~\bibnamefont
  {Garrod}}\ and\ \bibinfo {author} {\bibfnamefont {J.~K.}\ \bibnamefont
  {Percus}},\ }\bibfield  {title} {\bibinfo {title} {Reduction of the
  {$N$}-particle variational problem},\ }\href
  {https://doi.org/10.1063/1.1704098} {\bibfield  {journal} {\bibinfo
  {journal} {J. Math. Phys.}\ }\textbf {\bibinfo {volume} {5}},\ \bibinfo
  {pages} {1756} (\bibinfo {year} {1964})}\BibitemShut {NoStop}%
\bibitem [{\citenamefont {Mazziotti}(2006)}]{Mazziotti2006}%
  \BibitemOpen
  \bibfield  {author} {\bibinfo {author} {\bibfnamefont {D.~A.}\ \bibnamefont
  {Mazziotti}},\ }\bibfield  {title} {\bibinfo {title} {Variational
  reduced-density-matrix method using three-particle {$N$}-representability
  conditions with application to many-electron molecules},\ }\href
  {https://doi.org/10.1103/PhysRevA.74.032501} {\bibfield  {journal} {\bibinfo
  {journal} {Phys. Rev. A}\ }\textbf {\bibinfo {volume} {74}},\ \bibinfo
  {pages} {032501} (\bibinfo {year} {2006})}\BibitemShut {NoStop}%
\bibitem [{\citenamefont {Mazziotti}(2023)}]{Mazziotti2023}%
  \BibitemOpen
  \bibfield  {author} {\bibinfo {author} {\bibfnamefont {D.~A.}\ \bibnamefont
  {Mazziotti}},\ }\bibfield  {title} {\bibinfo {title} {Quantum many-body
  theory from a solution of the {$N$}-representability problem},\ }\href
  {https://doi.org/10.1103/PhysRevLett.130.153001} {\bibfield  {journal}
  {\bibinfo  {journal} {Phys. Rev. Lett.}\ }\textbf {\bibinfo {volume} {130}},\
  \bibinfo {pages} {153001} (\bibinfo {year} {2023})}\BibitemShut {NoStop}%
\bibitem [{\citenamefont {Mazziotti}(2026)}]{Mazziotti.2026}%
  \BibitemOpen
  \bibfield  {author} {\bibinfo {author} {\bibfnamefont {D.~A.}\ \bibnamefont
  {Mazziotti}},\ }\bibfield  {title} {\bibinfo {title} {{Representability for
  Quantum Theory beyond Particle-Number Conservation}},\ }\href
  {https://doi.org/10.1103/yt4s-dlwr} {\bibfield  {journal} {\bibinfo
  {journal} {Phys. Rev. Lett.}\ }\textbf {\bibinfo {volume} {136}},\ \bibinfo
  {pages} {120202} (\bibinfo {year} {2026})}\BibitemShut {NoStop}%
\bibitem [{\citenamefont {Recht}\ \emph {et~al.}(2010)\citenamefont {Recht},
  \citenamefont {Fazel},\ and\ \citenamefont {Parrilo}}]{Recht2010}%
  \BibitemOpen
  \bibfield  {author} {\bibinfo {author} {\bibfnamefont {B.}~\bibnamefont
  {Recht}}, \bibinfo {author} {\bibfnamefont {M.}~\bibnamefont {Fazel}},\ and\
  \bibinfo {author} {\bibfnamefont {P.~A.}\ \bibnamefont {Parrilo}},\
  }\bibfield  {title} {\bibinfo {title} {Guaranteed minimum-rank solutions of
  linear matrix equations via nuclear norm minimization},\ }\href
  {https://doi.org/10.1137/070697835} {\bibfield  {journal} {\bibinfo
  {journal} {SIAM Rev.}\ }\textbf {\bibinfo {volume} {52}},\ \bibinfo {pages}
  {471} (\bibinfo {year} {2010})}\BibitemShut {NoStop}%
\bibitem [{\citenamefont {Cand{\`e}s}\ and\ \citenamefont
  {Recht}(2009)}]{Candes2009}%
  \BibitemOpen
  \bibfield  {author} {\bibinfo {author} {\bibfnamefont {E.~J.}\ \bibnamefont
  {Cand{\`e}s}}\ and\ \bibinfo {author} {\bibfnamefont {B.}~\bibnamefont
  {Recht}},\ }\bibfield  {title} {\bibinfo {title} {Exact matrix completion via
  convex optimization},\ }\href {https://doi.org/10.1007/s10208-009-9045-5}
  {\bibfield  {journal} {\bibinfo  {journal} {Found. Comput. Math.}\ }\textbf
  {\bibinfo {volume} {9}},\ \bibinfo {pages} {717} (\bibinfo {year}
  {2009})}\BibitemShut {NoStop}%
\bibitem [{\citenamefont {Foley}\ and\ \citenamefont
  {Mazziotti}(2012)}]{Foley2012}%
  \BibitemOpen
  \bibfield  {author} {\bibinfo {author} {\bibfnamefont {J.~J.}\ \bibnamefont
  {Foley}}\ and\ \bibinfo {author} {\bibfnamefont {D.~A.}\ \bibnamefont
  {Mazziotti}},\ }\bibfield  {title} {\bibinfo {title} {Measurement-driven
  reconstruction of many-particle quantum processes by semidefinite programming
  with application to photosynthetic light harvesting},\ }\href
  {https://doi.org/10.1103/PhysRevA.86.012512} {\bibfield  {journal} {\bibinfo
  {journal} {Phys. Rev. A}\ }\textbf {\bibinfo {volume} {86}},\ \bibinfo
  {pages} {012512} (\bibinfo {year} {2012})}\BibitemShut {NoStop}%
\bibitem [{\citenamefont {Suhai}(1994)}]{Suhai1994}%
  \BibitemOpen
  \bibfield  {author} {\bibinfo {author} {\bibfnamefont {S.}~\bibnamefont
  {Suhai}},\ }\bibfield  {title} {\bibinfo {title} {Electron correlation in
  extended systems: {F}ourth-order many-body perturbation theory and
  density-functional methods applied to an infinite chain of hydrogen atoms},\
  }\href {https://doi.org/10.1103/PhysRevB.50.14791} {\bibfield  {journal}
  {\bibinfo  {journal} {Phys. Rev. B}\ }\textbf {\bibinfo {volume} {50}},\
  \bibinfo {pages} {14791} (\bibinfo {year} {1994})}\BibitemShut {NoStop}%
\bibitem [{\citenamefont {Hehre}\ \emph {et~al.}(1969)\citenamefont {Hehre},
  \citenamefont {Stewart},\ and\ \citenamefont {Pople}}]{Hehre1969}%
  \BibitemOpen
  \bibfield  {author} {\bibinfo {author} {\bibfnamefont {W.~J.}\ \bibnamefont
  {Hehre}}, \bibinfo {author} {\bibfnamefont {R.~F.}\ \bibnamefont {Stewart}},\
  and\ \bibinfo {author} {\bibfnamefont {J.~A.}\ \bibnamefont {Pople}},\
  }\bibfield  {title} {\bibinfo {title} {Self-consistent molecular-orbital
  methods. {I}. {U}se of {G}aussian expansions of {S}later-type atomic
  orbitals},\ }\href {https://doi.org/10.1063/1.1672392} {\bibfield  {journal}
  {\bibinfo  {journal} {J. Chem. Phys.}\ }\textbf {\bibinfo {volume} {51}},\
  \bibinfo {pages} {2657} (\bibinfo {year} {1969})}\BibitemShut {NoStop}%
\bibitem [{\citenamefont {Wang}\ and\ \citenamefont
  {Mazziotti}(2024)}]{Wang2024CI}%
  \BibitemOpen
  \bibfield  {author} {\bibinfo {author} {\bibfnamefont {Y.}~\bibnamefont
  {Wang}}\ and\ \bibinfo {author} {\bibfnamefont {D.~A.}\ \bibnamefont
  {Mazziotti}},\ }\bibfield  {title} {\bibinfo {title} {Quantum simulation of
  conical intersections},\ }\href {https://doi.org/10.1039/D4CP00391H}
  {\bibfield  {journal} {\bibinfo  {journal} {Phys. Chem. Chem. Phys.}\
  }\textbf {\bibinfo {volume} {26}},\ \bibinfo {pages} {11491} (\bibinfo {year}
  {2024})}\BibitemShut {NoStop}%
\bibitem [{\citenamefont {Wang}\ \emph
  {et~al.}(2025{\natexlab{b}})\citenamefont {Wang}, \citenamefont {Cianci},
  \citenamefont {Avdic}, \citenamefont {Dutta}, \citenamefont {Warren},
  \citenamefont {Allen}, \citenamefont {Vu}, \citenamefont {Santos},
  \citenamefont {Batista},\ and\ \citenamefont {Mazziotti}}]{Wang2025cytosine}%
  \BibitemOpen
  \bibfield  {author} {\bibinfo {author} {\bibfnamefont {Y.}~\bibnamefont
  {Wang}}, \bibinfo {author} {\bibfnamefont {C.}~\bibnamefont {Cianci}},
  \bibinfo {author} {\bibfnamefont {I.}~\bibnamefont {Avdic}}, \bibinfo
  {author} {\bibfnamefont {R.}~\bibnamefont {Dutta}}, \bibinfo {author}
  {\bibfnamefont {S.}~\bibnamefont {Warren}}, \bibinfo {author} {\bibfnamefont
  {B.}~\bibnamefont {Allen}}, \bibinfo {author} {\bibfnamefont {N.~P.}\
  \bibnamefont {Vu}}, \bibinfo {author} {\bibfnamefont {L.~F.}\ \bibnamefont
  {Santos}}, \bibinfo {author} {\bibfnamefont {V.~S.}\ \bibnamefont
  {Batista}},\ and\ \bibinfo {author} {\bibfnamefont {D.~A.}\ \bibnamefont
  {Mazziotti}},\ }\bibfield  {title} {\bibinfo {title} {Characterizing conical
  intersections of nucleobases on quantum computers},\ }\href
  {https://doi.org/10.1021/acs.jctc.4c01434} {\bibfield  {journal} {\bibinfo
  {journal} {J. Chem. Theory Comput.}\ }\textbf {\bibinfo {volume} {21}},\
  \bibinfo {pages} {1213} (\bibinfo {year} {2025}{\natexlab{b}})}\BibitemShut
  {NoStop}%
\bibitem [{\citenamefont {Pederzoli}\ \emph {et~al.}(2011)\citenamefont
  {Pederzoli}, \citenamefont {Pittner}, \citenamefont {Barbatti},\ and\
  \citenamefont {Lischka}}]{pederzoli2011nonadiabatic}%
  \BibitemOpen
  \bibfield  {author} {\bibinfo {author} {\bibfnamefont {M.}~\bibnamefont
  {Pederzoli}}, \bibinfo {author} {\bibfnamefont {J.}~\bibnamefont {Pittner}},
  \bibinfo {author} {\bibfnamefont {M.}~\bibnamefont {Barbatti}},\ and\
  \bibinfo {author} {\bibfnamefont {H.}~\bibnamefont {Lischka}},\ }\bibfield
  {title} {\bibinfo {title} {Nonadiabatic molecular dynamics study of the
  cis--trans photoisomerization of azobenzene excited to the {S}$_1$ state},\
  }\href {https://doi.org/10.1021/jp2013094} {\bibfield  {journal} {\bibinfo
  {journal} {J. Phys. Chem. A}\ }\textbf {\bibinfo {volume} {115}},\ \bibinfo
  {pages} {11136} (\bibinfo {year} {2011})}\BibitemShut {NoStop}%
\bibitem [{\citenamefont {Javadi-Abhari}\ \emph {et~al.}(2024)\citenamefont
  {Javadi-Abhari}, \citenamefont {Treinish}, \citenamefont {Krsulich},
  \citenamefont {Wood}, \citenamefont {Lishman}, \citenamefont {Gacon},
  \citenamefont {Martiel}, \citenamefont {Nation}, \citenamefont {Bishop},
  \citenamefont {Cross}, \citenamefont {Johnson},\ and\ \citenamefont
  {Gambetta}}]{Qiskit}%
  \BibitemOpen
  \bibfield  {author} {\bibinfo {author} {\bibfnamefont {A.}~\bibnamefont
  {Javadi-Abhari}}, \bibinfo {author} {\bibfnamefont {M.}~\bibnamefont
  {Treinish}}, \bibinfo {author} {\bibfnamefont {K.}~\bibnamefont {Krsulich}},
  \bibinfo {author} {\bibfnamefont {C.~J.}\ \bibnamefont {Wood}}, \bibinfo
  {author} {\bibfnamefont {J.}~\bibnamefont {Lishman}}, \bibinfo {author}
  {\bibfnamefont {J.}~\bibnamefont {Gacon}}, \bibinfo {author} {\bibfnamefont
  {S.}~\bibnamefont {Martiel}}, \bibinfo {author} {\bibfnamefont {P.~D.}\
  \bibnamefont {Nation}}, \bibinfo {author} {\bibfnamefont {L.~S.}\
  \bibnamefont {Bishop}}, \bibinfo {author} {\bibfnamefont {A.~W.}\
  \bibnamefont {Cross}}, \bibinfo {author} {\bibfnamefont {B.~R.}\ \bibnamefont
  {Johnson}},\ and\ \bibinfo {author} {\bibfnamefont {J.~M.}\ \bibnamefont
  {Gambetta}},\ }\bibfield  {title} {\bibinfo {title} {Quantum computing with
  {Q}iskit},\ }\bibfield  {journal} {\bibinfo  {journal} {arXiv preprint
  arXiv:2405.08810}\ }\href {https://doi.org/10.48550/arXiv.2405.08810}
  {10.48550/arXiv.2405.08810} (\bibinfo {year} {2024})\BibitemShut {NoStop}%
\bibitem [{\citenamefont {Sung}\ \emph {et~al.}(2026)\citenamefont {Sung},
  \citenamefont {Choi}, \citenamefont {Amico}, \citenamefont {Andrews},
  \citenamefont {Ayantuna}, \citenamefont {Kawashima}, \citenamefont {Lin},
  \citenamefont {Omanovic}, \citenamefont {Piccinelli}, \citenamefont
  {Robledo~Moreno}, \citenamefont {Saki}, \citenamefont {Shee}, \citenamefont
  {Shin}, \citenamefont {Tran}, \citenamefont {Ueda}, \citenamefont {Zhang},\
  and\ \citenamefont {Motta}}]{ffsim}%
  \BibitemOpen
  \bibfield  {author} {\bibinfo {author} {\bibfnamefont {K.~J.}\ \bibnamefont
  {Sung}}, \bibinfo {author} {\bibfnamefont {I.}~\bibnamefont {Choi}}, \bibinfo
  {author} {\bibfnamefont {M.}~\bibnamefont {Amico}}, \bibinfo {author}
  {\bibfnamefont {B.}~\bibnamefont {Andrews}}, \bibinfo {author} {\bibfnamefont
  {E.}~\bibnamefont {Ayantuna}}, \bibinfo {author} {\bibfnamefont
  {Y.}~\bibnamefont {Kawashima}}, \bibinfo {author} {\bibfnamefont {W.-H.}\
  \bibnamefont {Lin}}, \bibinfo {author} {\bibfnamefont {D.}~\bibnamefont
  {Omanovic}}, \bibinfo {author} {\bibfnamefont {S.}~\bibnamefont
  {Piccinelli}}, \bibinfo {author} {\bibfnamefont {J.}~\bibnamefont
  {Robledo~Moreno}}, \bibinfo {author} {\bibfnamefont {A.~A.}\ \bibnamefont
  {Saki}}, \bibinfo {author} {\bibfnamefont {J.}~\bibnamefont {Shee}}, \bibinfo
  {author} {\bibfnamefont {S.}~\bibnamefont {Shin}}, \bibinfo {author}
  {\bibfnamefont {M.~C.}\ \bibnamefont {Tran}}, \bibinfo {author}
  {\bibfnamefont {K.}~\bibnamefont {Ueda}}, \bibinfo {author} {\bibfnamefont
  {H.}~\bibnamefont {Zhang}},\ and\ \bibinfo {author} {\bibfnamefont
  {M.}~\bibnamefont {Motta}},\ }\bibfield  {title} {\bibinfo {title} {ffsim:
  {F}aster simulation of fermionic quantum circuits},\ }\bibfield  {journal}
  {\bibinfo  {journal} {arXiv preprint arXiv:2605.03123}\ }\href
  {https://doi.org/10.48550/arXiv.2605.03123} {10.48550/arXiv.2605.03123}
  (\bibinfo {year} {2026})\BibitemShut {NoStop}%
\bibitem [{map(2026)}]{maple_2025}%
  \BibitemOpen
  \href {https://www.maplesoft.com} {\emph {\bibinfo {title} {Maple}}}\
  (\bibinfo  {publisher} {Maplesoft, Waterloo},\ \bibinfo {year}
  {2026})\BibitemShut {NoStop}%
\bibitem [{rdm(2026)}]{rdmchem_2025}%
  \BibitemOpen
  \href {https://www.rdmchem.com} {\emph {\bibinfo {title} {RDMChem, Quantum
  Chemistry Package}}}\ (\bibinfo  {publisher} {Maplesoft, Waterloo},\ \bibinfo
  {year} {2026})\BibitemShut {NoStop}%
\bibitem [{\citenamefont {Mazziotti}(2011)}]{M2011}%
  \BibitemOpen
  \bibfield  {author} {\bibinfo {author} {\bibfnamefont {D.~A.}\ \bibnamefont
  {Mazziotti}},\ }\bibfield  {title} {\bibinfo {title} {Large-scale
  semidefinite programming for many-electron quantum mechanics},\ }\href
  {https://doi.org/10.1103/PhysRevLett.106.083001} {\bibfield  {journal}
  {\bibinfo  {journal} {Phys. Rev. Lett.}\ }\textbf {\bibinfo {volume} {106}},\
  \bibinfo {pages} {083001} (\bibinfo {year} {2011})}\BibitemShut {NoStop}%
\bibitem [{\citenamefont {Beharry}\ and\ \citenamefont
  {Woolley}(2011)}]{Beharry2011}%
  \BibitemOpen
  \bibfield  {author} {\bibinfo {author} {\bibfnamefont {A.~A.}\ \bibnamefont
  {Beharry}}\ and\ \bibinfo {author} {\bibfnamefont {G.~A.}\ \bibnamefont
  {Woolley}},\ }\bibfield  {title} {\bibinfo {title} {Azobenzene photoswitches
  for biomolecules},\ }\href {https://doi.org/10.1039/C1CS15023E} {\bibfield
  {journal} {\bibinfo  {journal} {Chem. Soc. Rev.}\ }\textbf {\bibinfo {volume}
  {40}},\ \bibinfo {pages} {4422} (\bibinfo {year} {2011})}\BibitemShut
  {NoStop}%
\bibitem [{\citenamefont {García-Amorós}\ and\ \citenamefont
  {Velasco}(2012)}]{Garcia2012}%
  \BibitemOpen
  \bibfield  {author} {\bibinfo {author} {\bibfnamefont {J.}~\bibnamefont
  {García-Amorós}}\ and\ \bibinfo {author} {\bibfnamefont {D.}~\bibnamefont
  {Velasco}},\ }\bibfield  {title} {\bibinfo {title} {Recent advances towards
  azobenzene-based light-driven real-time information-transmitting materials},\
  }\href {https://doi.org/10.3762/bjoc.8.113} {\bibfield  {journal} {\bibinfo
  {journal} {Beilstein J. Org. Chem.}\ }\textbf {\bibinfo {volume} {8}},\
  \bibinfo {pages} {1003} (\bibinfo {year} {2012})}\BibitemShut {NoStop}%
\bibitem [{\citenamefont {Kang}\ and\ \citenamefont {Yang}(2022)}]{Kang2022}%
  \BibitemOpen
  \bibfield  {author} {\bibinfo {author} {\bibfnamefont {H.~S.}\ \bibnamefont
  {Kang}}\ and\ \bibinfo {author} {\bibfnamefont {S.}~\bibnamefont {Yang}},\
  }\bibfield  {title} {\bibinfo {title} {Photopatterning via photofluidization
  of azobenzene polymers},\ }\href {https://doi.org/10.37188/lam.2022.003}
  {\bibfield  {journal} {\bibinfo  {journal} {Light Adv. Manuf.}\ }\textbf
  {\bibinfo {volume} {3}},\ \bibinfo {pages} {1} (\bibinfo {year}
  {2022})}\BibitemShut {NoStop}%
\bibitem [{\citenamefont {Merino}\ and\ \citenamefont
  {Ribagorda}(2012)}]{Merino2012}%
  \BibitemOpen
  \bibfield  {author} {\bibinfo {author} {\bibfnamefont {E.}~\bibnamefont
  {Merino}}\ and\ \bibinfo {author} {\bibfnamefont {M.}~\bibnamefont
  {Ribagorda}},\ }\bibfield  {title} {\bibinfo {title} {Control over molecular
  motion using the cis–trans photoisomerization of the azo group},\ }\href
  {https://doi.org/10.3762/bjoc.8.119} {\bibfield  {journal} {\bibinfo
  {journal} {Beilstein J. Org. Chem.}\ }\textbf {\bibinfo {volume} {8}},\
  \bibinfo {pages} {1071} (\bibinfo {year} {2012})}\BibitemShut {NoStop}%
\bibitem [{\citenamefont {Pace}\ \emph {et~al.}(2007)\citenamefont {Pace},
  \citenamefont {Ferri}, \citenamefont {Grave}, \citenamefont {Elbing},
  \citenamefont {von Hänisch}, \citenamefont {Zharnikov}, \citenamefont
  {Mayor}, \citenamefont {Rampi},\ and\ \citenamefont {Samorì}}]{Pace2007}%
  \BibitemOpen
  \bibfield  {author} {\bibinfo {author} {\bibfnamefont {G.}~\bibnamefont
  {Pace}}, \bibinfo {author} {\bibfnamefont {V.}~\bibnamefont {Ferri}},
  \bibinfo {author} {\bibfnamefont {C.}~\bibnamefont {Grave}}, \bibinfo
  {author} {\bibfnamefont {M.}~\bibnamefont {Elbing}}, \bibinfo {author}
  {\bibfnamefont {C.}~\bibnamefont {von Hänisch}}, \bibinfo {author}
  {\bibfnamefont {M.}~\bibnamefont {Zharnikov}}, \bibinfo {author}
  {\bibfnamefont {M.}~\bibnamefont {Mayor}}, \bibinfo {author} {\bibfnamefont
  {M.~A.}\ \bibnamefont {Rampi}},\ and\ \bibinfo {author} {\bibfnamefont
  {P.}~\bibnamefont {Samorì}},\ }\bibfield  {title} {\bibinfo {title}
  {Cooperative light-induced molecular movements of highly ordered azobenzene
  self-assembled monolayers},\ }\href {https://doi.org/10.1073/pnas.0703748104}
  {\bibfield  {journal} {\bibinfo  {journal} {Proc. Natl. Acad. Sci. U.S.A.}\
  }\textbf {\bibinfo {volume} {104}},\ \bibinfo {pages} {9937} (\bibinfo {year}
  {2007})}\BibitemShut {NoStop}%
\bibitem [{\citenamefont {Szymanski}\ \emph {et~al.}(2013)\citenamefont
  {Szymanski}, \citenamefont {Beierle}, \citenamefont {Kistemaker},
  \citenamefont {Velema},\ and\ \citenamefont {Feringa}}]{Szymanski2013}%
  \BibitemOpen
  \bibfield  {author} {\bibinfo {author} {\bibfnamefont {W.}~\bibnamefont
  {Szymanski}}, \bibinfo {author} {\bibfnamefont {J.~M.}\ \bibnamefont
  {Beierle}}, \bibinfo {author} {\bibfnamefont {H.~A.~V.}\ \bibnamefont
  {Kistemaker}}, \bibinfo {author} {\bibfnamefont {W.~A.}\ \bibnamefont
  {Velema}},\ and\ \bibinfo {author} {\bibfnamefont {B.~L.}\ \bibnamefont
  {Feringa}},\ }\bibfield  {title} {\bibinfo {title} {Reversible photocontrol
  of biological systems by the incorporation of molecular photoswitches},\
  }\href {https://doi.org/10.1021/cr300179f} {\bibfield  {journal} {\bibinfo
  {journal} {Chem. Rev.}\ }\textbf {\bibinfo {volume} {113}},\ \bibinfo {pages}
  {6114} (\bibinfo {year} {2013})}\BibitemShut {NoStop}%
\bibitem [{\citenamefont {Wang}\ and\ \citenamefont
  {Mazziotti}(2026)}]{WangMazziottiGitHub}%
  \BibitemOpen
  \bibfield  {author} {\bibinfo {author} {\bibfnamefont {Y.}~\bibnamefont
  {Wang}}\ and\ \bibinfo {author} {\bibfnamefont {D.~A.}\ \bibnamefont
  {Mazziotti}},\ }\href
  {https://github.com/damazz/OptimallyWeightedConstrainedShadowTomography}
  {\bibinfo {title} {Optimally weighted constrained shadow tomography}},\
  \bibinfo {howpublished} {GitHub repository} (\bibinfo {year}
  {2026})\BibitemShut {NoStop}%
\end{thebibliography}%

\end{document}